\documentclass[floatfix,preprint,showpacs,showemail,preprintnumbers,amsfonts,pre]{revtex4}
\usepackage{graphicx}
\usepackage{bm}

\begin{document}
\title{
An individual-based predator-prey model for biological coevolution:
Fluctuations, stability, and community structure 
}
\author{Per Arne Rikvold$^{1,2}$}\email{rikvold@scs.fsu.edu}
\author{Volkan Sevim$^{1}$}\email{sevim@scs.fsu.edu}
\affiliation{
$^1$School of Computational Science, 
Center for Materials Research and Technology, and Department of Physics,
Florida State University, Tallahassee, Florida 32306-4120, USA\\
$^2$National High Magnetic Field Laboratory, 
Tallahassee, Florida, 32310-3706, USA\\
}
\date{\today}

\begin{abstract}
We study an individual-based predator-prey model of biological
coevolution, using linear stability analysis and large-scale kinetic
Monte Carlo simulations. The model exhibits approximate $1/f$ noise in
diversity and population-size fluctuations, and it generates a sequence
of quasi-steady communities in the form of simple food webs. These
communities are quite resilient toward the loss of one or a few species,
which is reflected in different power-law exponents for the durations of
communities and the lifetimes of species. The exponent for the former is
near $-1$, while the latter is close to $-2$. 
Statistical characteristics of the evolving communities, including
degree (predator and prey) distributions and proportions of basal,
intermediate, and top species, compare reasonably with data
for real food webs. 
\end{abstract}


\maketitle

\section{Introduction}
\label{sec:Int}

Biological evolution presents many problems concerning
interacting multi-entity systems far from equilibrium that are well suited
for methods from nonequilibrium
statistical physics \cite{BAAK99,DROS01}. Among these are questions
concerning the dynamics of the emergence and extinction of species on
macroevolutionary timescales \cite{NEWM99B,NEWM03,BORN06}.  
Traditionally it has been common to treat ecological and evolutionary
processes on very different timescales. However, it has recently been
realized that evolution often can take place on short timescales,
comparable to those of ecological processes
\cite{THOM98,THOM99,DROS01B,YOSH03}. 
A well-known example of very rapid evolution is provided by the cichlid
fishes of East Africa \cite{KOCH04}. 
Several models have therefore been proposed that, 
while spanning disparate scales of temporal and taxonomic resolution, 
consider the complex problem of coevolution of species in a fitness
landscape that constantly changes with the composition of the community.
Early contributions were simulations of parapatric and sympatric speciation
\cite{CROS70} and the coupled $NK$ model with
population dynamics \cite{KAUF91,KAUF93}.
More recent work includes the Webworld model
\cite{CALD98,DROS01B,DROS04},
the tangled-nature model
\cite{CHRI02,HALL02,COLL03}
and simplified versions of the latter
\cite{RIKV03,ZIA04,SEVI06},
as well as network models
\cite{CHOW03A,CHOW05}.
Recently, large individual-based simulations have also been performed of
parapatric and sympatric speciation \cite{GAVR98,GAVR00} and of
adaptive radiation \cite{GAVR05}.

Many of the models discussed above are deliberately quite simple, aiming to
elucidate universal features that are largely independent of the
finer details of the ecological interactions and the evolutionary
mechanisms. Such 
features may include lifetime distributions for species and communities,
as well as other aspects of extinction statistics,
statistical properties of fluctuations in diversity and population
sizes, and the structure and dynamics of food webs that develop and change
with time. 

In the present paper we continue our study of a simplified version of the
tangled-nature model. In the early studies of this individual-based model 
of coevolution 
\cite{RIKV03,ZIA04,SEVI06}, the interspecies interactions, which are
described by an interaction matrix $\bf M$ \cite{SOLE96B}, were random and
could produce any combination of pair interactions: favorable-favorable, 
deleterious-deleterious, or favorable-deleterious. 
Under those conditions the model was found to
evolve through a sequence of quasi-stable communities,  
in which all species interact 
with mutually favorable interactions, i.e., mutualistic or
symbiotic communities. For that reason we shall hereafter refer to that
version of the 
model as {\it the mutualistic model\/}. Here we instead concentrate on a
version that specifically describes the evolution of 
predator-prey communities. This restriction is enforced by means of an
antisymmetric interaction matrix, so that an interaction that is
favorable for one member of a pair is deleterious for the other, and 
{\it vice versa\/}. Many aspects of the dynamics of
this predator-prey model are similar to the mutualistic model, such as
approximate 
$1/f$ noise in species diversity and population sizes, and power-law
distributions of the durations of communities and the lifetimes of
species. However, some of the power-law exponents are different, and,
most importantly, the predator-prey model produces communities that take
the form of simple food webs. 
It also shares with the mutualistic model the property that mean
population sizes and stability properties of fixed-point communities can be
calculated exactly in the absence of mutations. 
Comparisons of some aspects of the predator-prey model (with
a much smaller number of potential species than used here) to  
those of the mutualistic model have been 
presented in Refs.~\cite{RIKV05A,RIKV06}. 
The focus of the present paper is a much more detailed discussion
of the dynamics and the structure of the resulting food webs 
(including comparison with real food webs) for 
the predator-prey model with a large number of
potential species. For this purpose we use both exact linear-stability
analysis and large-scale kinetic Monte Carlo simulations.  
In particular we wish to study the fluctuations in the statistically
stationary state that develops for long times. A motivation is the hope
that understanding of these stationary-state fluctuations can provide
information about the system's sensitivity to external perturbations in
a way analogous to a fluctuation-dissipation relation \cite{SATO03}. 
We therefore carry out very long simulations. 

The rest of this paper is organized as follows. The model is 
presented in Sec.~\ref{sec:Mod}. Exact
linear-stability analysis is performed in Sec.~\ref{sec:LSA}, including
fixed-point population sizes in Sec.~\ref{sec:fpc} and stability
considerations in Sec.~\ref{sec:stab}. Numerical results 
are presented in Sec.~\ref{sec:NR},
including species abundance distributions and
time series of diversities and population sizes
(Sec.~\ref{sec:TS}), power spectral densities (Sec.~\ref{sec:PSD}),
species lifetimes (Sec.~\ref{sec:life}), 
durations of evolutionarily quiet and active periods (Sec.~\ref{sec:Q}),
and community structure and
stability with comparison with real food webs
(Sec.~\ref{sec:web}). Our conclusions are summarized in 
Sec.~\ref{sec:Concl}, and the method used to calculate the interaction
matrices for systems with a large number of potential species is
explained in Appendix~\ref{sec:AA}.

\section{Model}
\label{sec:Mod}

The model considered here is a version of the macroevolution model
introduced by Rikvold and Zia \cite{RIKV03} as a simplification of
the tangled-nature model of Jensen and coworkers \cite{HALL02,CHRI02,COLL03}. 
In this version, the interspecies interactions are constrained to
represent a pure predator-prey system. 
As in Ref.~\cite{RIKV03}, 
selection is provided by the reproduction rates in an individual-based,
simplified multispecies population-dynamics model with nonoverlapping
generations. This interacting birth/death
process is augmented to enable evolution of new species by a mutation
mechanism.
The mutations act on a haploid, binary ``genome" of length $L$, as
introduced by Eigen for molecular evolution
\cite{EIGE71,EIGE88}. This bit string defines the species, which
are identified by the integer label $I \in [0,2^L-1]$.
Typically, only a few of these $2^L$ potential species
are resident in the community at any one time.

During reproduction, an offspring individual may undergo a mutation 
that flips a randomly chosen gene ($0 \rightarrow 1$ or $1 \rightarrow 0$) 
with a small probability, $\mu$. 
The mutation thus corresponds to diffusional moves from corner to
corner along the edges of an $L$-dimensional hypercube
\cite{GAVR99,GAVR04}.  A mutated individual
is assumed to belong to a different species than its parent, with
different properties. Genotype and phenotype are thus in one-to-one
correspondence in this model.
This is clearly a highly idealized picture, and it is introduced
to maximize the pool of different species available within the
computational resources. This picture is justified by
a large-scale computational study of
the mutualistic version of the model studied in Ref.~\cite{RIKV03}, 
in which species that differ by as many as
$L/2$ bits have correlated properties \cite{SEVI06}.
Remarkably, this study reveals
that the more realistic, correlated model has
long-time dynamical properties very similar to the uncorrelated model.

The reproduction probability $P_I(t)$ for an individual of species $I$
in generation $t$ depends on the individual's ability to
utilize the amount $R$ of available external resources,
and on its interactions with the population sizes $n_J(t)$
of all the species present in the community at that time.
The dependence of $P_I$ on the set of $n_J$ is determined by an
{\it interaction matrix\/} $\bf M$ \cite{SOLE96B}
with offdiagonal elements ${M}_{IJ}$ that are continuously and symmetrically
distributed on the interval $[-1,+1]$ in a
way defined specifically in the next paragraph.
The elements of $\bf M$ are chosen randomly at the beginning of each
simulation run and are subsequently kept constant throughout
the run (quenched randomness). (For a discussion of how the
matrix elements are created for $L > 13$, in which case the 
$2^L \times 2^L$ matrix does 
not fit into the memory of a standard workstation, see
Appendix~\ref{sec:AA}. This method leads to a distribution that is
triangular on $[-1,+1]$.)

In contrast to our previously studied model
\cite{RIKV03,ZIA04,SEVI06}, in which the interaction matrix has no particular
structure, the predator-prey dynamics is enforced by the
requirement that the off-diagonal part of 
$\bf M$ must be antisymmetric. Thus, if $M_{IJ} >
0$ and $M_{JI} < 0$, then species $I$ is the predator and $J$ the prey, 
and {\it vice versa\/}.
In order to keep the connectance of the resulting communities consistent
with food webs observed in nature \cite{DUNN02,GARL04} the
$(M_{IJ},M_{JI})$ pairs are chosen nonzero with probability $c=0.1$. 
The nonzero elements in the upper triangle of $\bf M$ are chosen
independently from the triangular distribution on $[-1,+1]$, described
in Appendix~\ref{sec:AA}. Self-competition is included in the model by
choosing the diagonal elements of $M$ randomly and uniformly from 
$[-1,0)$.  

The reproduction probability for species $I$, $P_I(t)$,
depends on $R$ and the set $\{n_J(t)\}$ through the nonlinear form,
\begin{equation}
P_I(t) = \frac{1}{1 + \exp[-\Delta_I(R,\{n_J(t)\})]} \;,
\label{eq:PI}
\end{equation}
where
\begin{equation}
\Delta_I(R,\{n_J(t)\}) = - b_I + \eta_I R / N_{\rm tot} 
+ \sum_J M_{IJ} n_J(t) / N_{\rm tot} 
\;.
\label{eq:Delta}
\end{equation}
Here $b_I$ is the ``cost" of reproduction for species $I$ (always positive), 
and $\eta_I$ (positive for primary producers or autotrophs,
and zero for consumers or heterotrophs) is the ability of
individuals of species $I$ to utilize the external resource $R$.
The latter 
is renewed at the same level every generation and does {\it not\/} have
independent dynamics. 
The total population size is $N_{\rm tot}(t) = \sum_J n_J(t)$.
[In contrast, the total number of {\it species\/} present in
generation $t$
(the species richness) will be defined as $\mathcal{N}(t)$.]
The population-limiting
reproduction costs $b_I$ are chosen randomly and uniformly from
the interval $(0,+1]$. 
Only a proportion $p$ of the $2^L$ potential species are producers that
can directly utilize the resource. (For the numerical data shown here,
we use $p=0.05$.) Thus, with probability $(1-p)$ the resource coupling
$\eta_I=0$, representing consumers, while with probability $p$ the
$\eta_I$ are independently and uniformly distributed on $(0,+1]$,
representing producers of varying efficiency. 
In addition to the constraints on $\bf M$ mentioned above, we require
that producers always are the prey of consumers. Thus, the case 
$\eta_I >0$ and $\eta_J=0$ with $M_{IJ} \equiv -M_{JI} >0$ is forbidden
and is changed during setup of the matrix by reversing the signs of the
interactions for the pair in question. 

For large positive $\Delta_I$,
(small birth cost, strong coupling to the external resources,
and more prey than predators),
the individual almost certainly reproduces, giving rise to $F$
offspring. In the opposite limit of large negative $\Delta_I$,
(large birth cost, weak or no coupling to the external resources,
and/or more predators than prey),
it almost certainly dies without offspring.
The nonlinear dependence of $P_I$ on $\Delta_I$
thus limits the growth rate of the population size, even
under extremely favorable conditions. It also sets a
practical negative limit on $\Delta_I$,
below which conditions are so unfavorable that reproduction
is virtually impossible. 
(A more general version of Eq.~(\ref{eq:Delta}), in which population
growth is directly limited by a ``Verhulst factor" \cite{VERH1838} or
``environmental carrying capacity" \cite{MURR89} 
as is necessary in models that allow mutualistic interactions, 
is discussed in Ref.~\cite{RIKV06}.) 
We note that ``energy dissipation" in this model is achieved through the
birth costs $b_I$ and the self-competition terms $M_{II}$. Equivalent
effects could have been produced by making the positive $M_{IJ}$ smaller
than the corresponding negative ones by an ``ecological efficiency" factor 
between zero and unity, as in the Webworld model \cite{DROS01B,CALD98,DROS04}. 

The normalization of $\Delta_I$ with $N_{\rm tot}$ implies 
global competition. This is not very realistic, but it
enables us to find exact expressions for the stationary values of
the average population sizes in the mutation-free limit. (See
Sec.~\ref{sec:fpc}.) The model
can thus be used as a benchmark for more realistic ones in future research. 


An analytic approximation describing the development in time of the
mean population sizes (averaged over independent
realizations), $\langle n_I(t) \rangle$, can be
written as a set of coupled difference equations,
\begin{eqnarray}
\langle n_I(t+1) \rangle
&=& \langle n_I(t) \rangle FP_I(R,\{\langle n_J(t)\rangle \})[1-\mu ]
\nonumber\\
&& +(\mu/L)F\sum_{K(I)}\langle n_{K(I)}(t)
\rangle P_{K(I)}(R,\{\langle n_J(t) \rangle \}) 
\;,
\label{eq:MF}
\end{eqnarray}
where $K(I)$ is the set of species that can be generated from species $I$
by a single mutation (``nearest neighbors" of $I$ in genotype space).

\section{Linear Stability Analysis}
\label{sec:LSA}

\subsection{Fixed-point communities}
\label{sec:fpc}

An advantage of the model studied here is that its fixed-point
communities in the mutation-free limit can be found exactly within
a mean-field approximation based on Eq.~(\ref{eq:MF}) \cite{RIKV03}. 
To obtain a stationary solution 
for a community of $\mathcal{N}$ species, we must require $P_I=1/F$
for all $\mathcal{N}$ species.
Equations (\ref{eq:PI}) and~(\ref{eq:Delta}) then give rise to 
$\mathcal{N}$ linear relations, which can be written on the matrix
form 
\begin{equation}
- | \tilde{b} \rangle N_{\rm tot}^*
+ | \eta \rangle R
+ {\bf \hat{M}} | n^* \rangle
= 0
\;,
\label{eq:ss2}
\end{equation}
where $\tilde{b}_I = b_I - \ln(F-1)$, 
$| \tilde{b} \rangle$, $| \eta \rangle$, and $| n^* \rangle$ 
are the column vectors of $\tilde{b}_I$, $\eta_I$, and $n_I^*$, 
respectively (in all cases including only those $\mathcal{N}$
species that have nonzero $n_I^*$),
and $\bf \hat{M}$ is the corresponding $\mathcal{N}
\times \mathcal{N}$ submatrix of $\bf M$. (For simplicity, we
drop the $\langle \; \rangle$ notation for the average
population sizes, and the asterisk superscripts denote fixed-point
solutions.) 

The solution for $| n^* \rangle$ is
\begin{equation}
| n^* \rangle = - {\bf \hat{M}}^{-1} \left[ | \eta \rangle R
- | \tilde{b} \rangle N_{\rm tot}^*
\right]
\;,
\label{eq:ssn}
\end{equation}
where ${\bf \hat{M}}^{-1}$ is the inverse of ${\bf \hat{M}}$.
To find each $n_I^*$, we must first obtain
$N_{\rm tot}^* \equiv \langle 1 | n^* \rangle$, where 
$\langle 1 |$ is an $\mathcal{N}$-dimensional
row vector composed entirely of ones.
Multiplying Eq.~(\ref{eq:ssn}) from the left by $\langle 1 |$, we obtain
\begin{equation}
R \mathcal{E} + \Theta N_{\rm tot}^* = 0
\;,
\label{eq:quad}
\end{equation}
where the coefficients 
\begin{eqnarray}
\Theta =
\frac{1 - \langle 1 | {\bf \hat{M}}^{-1} | \tilde{b} \rangle}
     {\langle 1 | {\bf \hat{M}}^{-1} | 1 \rangle }
& \;\; \mbox{\rm and} \;\; &
\mathcal{E} = \frac{\langle 1 | {\bf \hat{M}}^{-1} | \eta \rangle}
                   {\langle 1 | {\bf \hat{M}}^{-1} | 1 \rangle }
\label{eq:quad2}
\end{eqnarray}
have been written with ${\langle 1 | {\bf \hat{M}}^{-1} | 1 \rangle }$ 
in the denominators in order to remain finite even for near-singular $\bf M$ 
\cite{RIKV06}. 
They can be viewed as an effective interaction strength and an effective
coupling to the external resource, respectively. 
The solution of Eq.~(\ref{eq:quad}) is 
\begin{equation}
N_{\rm tot}^* 
=
- \frac{R \mathcal{E}}{\Theta}
=
\frac{R \langle 1 | {\bf \hat{M}}^{-1} | \eta \rangle}
     {\langle 1 | {\bf \hat{M}}^{-1} | \tilde{b} \rangle - 1}
\;.
\label{eq:ntot}
\end{equation}
To find each $n_I^*$ separately, we now only need to insert this 
solution for $N_{\rm tot}^*$ in Eq.~(\ref{eq:ssn}).

Only those $| n^* \rangle$ that have all positive elements can represent
a {\it feasible\/} community \cite{ROBE74}.
If ${\bf \hat{M}} = {\bf 0}$ or is otherwise singular,
the set of equations (\ref{eq:ss2}) is
inconsistent for $\mathcal{N} >1$, unless $\tilde{b}_I$ and $\eta_I$
both are independent of $I$ (this case is equivalent to
$\mathcal{N}=1$).
The only possible stationary community then consists of one
single species, the one with the largest value of $\eta_I/\tilde{b}_I$.
This is a trivial example of competitive
exclusion \cite{HARD60,ARMS80,DENB86}.
If $\eta_I/\tilde{b}_I$ has the same value for all $\mathcal{N}$ values
of $I$, we have an example of a neutral model \cite{HUBB01}.

In Ref.~\cite{RIKV06} it was shown that Eq.~(\ref{eq:quad}) 
for fixed $\mathcal{E}$ and $\Theta$ can be seen as a maximization condition
for a ``community fitness" function,
\begin{equation}
\Phi(N_{\rm tot})
= \left( 1 - \frac{1}{F} \right)
\left(
R \mathcal{E} N_{\rm tot}
+ \frac{\Theta}{2} N_{\rm tot}^2
\right)
\;.
\label{eq:fit}
\end{equation}
This result would not be particularly remarkable if $\mathcal{E}$ and
$\Theta$ were externally fixed parameters. However, 
extinctions and mutations provide
a mechanism for both parameters to change as old species go extinct and
new species emerge. In our numerical simulations we find that their values
evolve toward and then fluctuate around values that
maximize $\Phi$, limited only by the internal
constraints on $\bf M$ and $| \tilde{b}_I \rangle$. In particular, this
means that $\Theta$ approaches closely to 0 from the negative side
\cite{RIKV06}. In the simulations presented in this paper 
(see Sec.~\ref{sec:NR}) averages over the final communities of twelve
independent simulation runs yield  $\overline{\Theta} \approx -0.10\pm0.03$ and 
$\overline{\mathcal{E}} \approx 0.7\pm0.2$, 
corresponding to $\overline{N_{\rm tot}^*} \approx (7.4\pm1.4) R$. 
In comparison, for fourteen random, feasible communities obtained as described
in Sec.~\ref{sec:webA} below, the corresponding parameters are 
$\overline{\Theta} \approx -0.39\pm0.06$ and 
$\overline{\mathcal{E}} \approx 0.66\pm0.08$,
corresponding to $\overline{N_{\rm tot}^*} \approx (2.0\pm0.3) R$. 
A detailed discussion of the statistical properties of these and other
quantities characteristic of the simulated communities are given in 
Sec.~\ref{sec:webB}. 

\subsection{Stability of fixed-point communities}
\label{sec:stab}

The internal stability of an $\mathcal{N}$-species
fixed-point community is obtained from the matrix of partial derivatives,
\begin{equation}
\left. \frac{\partial n_I(t+1)}{\partial n_J(t)} \right|_{| n^*
\rangle}
=
\delta_{IJ} + \Lambda_{IJ} \;,
\label{eq:stab}
\end{equation}
where $\delta_{IJ}$ is the Kronecker delta function and
$\Lambda_{IJ}$ are elements of the {\it community matrix\/}
$\bf \Lambda$ \citep{MURR89}. Straightforward differentiation
yields \cite{RIKV06}
\begin{equation}
\Lambda_{IJ} = \left( 1 - \frac{1}{F} \right) \frac{n_I^*}{N_{\rm
tot}^*}
\left[ M_{IJ}
- \frac{R \eta_I +({\bf \hat{M}}|n^*\rangle)_I}{N_{\rm tot}^*}
\right]
\;,
\label{eq:lam}
\end{equation}
where $({\bf \hat{M}}|n^*\rangle)_I$ is the element of the
column vector ${\bf \hat{M}}|n^*\rangle$, corresponding to species $I$.
In order for deviations from
the fixed point to decay monotonically in magnitude,
the magnitudes of the eigenvalues of the matrix of partial
derivatives
in Eq.~(\ref{eq:stab}), $\bf{\Lambda} + {\bf 1}$, where $\bf 1$ is
the
$\mathcal{N}$-dimensional unit matrix, must be less than unity.
The value of the fecundity used in this work, $F=2$, 
was chosen to satisfy this requirement for $\mathcal{N}=1$.

Since new species are created by mutations, we must also study the
stability of the fixed-point community toward ``invaders."
Consider a mutant invader $i$.
Then its multiplication rate, in the limit that
$n_i \ll n_J$ for all $\mathcal{N}$
species $J$ in the resident community, is given by
\begin{equation}
\frac{n_i(t+1)}{n_i(t)}
=
\frac{F}{1 + \exp \left[ -\Delta_i(R,\{n_J^*\}) \right]}
\;.
\label{eq:inv}
\end{equation}
The Lyapunov exponent, $\ln[ n_i(t+1)/n_i(t) ]$, is the {\it
invasion fitness\/} of the mutant with respect to the resident community
\citep{METZ92,DOEB00}. It will be studied numerically in
Sec.~\ref{sec:webA}.

\section{Numerical Results}
\label{sec:NR}

We performed twelve independent, 
long simulation runs of $2^{25} = 33\,554\,432$
generations of the model with the following parameters: 
genome length $L=20$ (1\,048\,576 potential species), 
external resource $R=2000$, and mutation rate $\mu = 10^{-3}$,
with connectance parameter $c = 0.1$ and a proportion $p=0.05$ of the
potential species as producers. 
These parameters were chosen to represent the realistic situation that
the number of species resident in the community at any time
is much smaller than the number of potential species
(i.e., that $\mathcal{N}(t) \ll 2^L$), and also that
$\mathcal{N}(t) \ll N_{\rm tot}(t)$ so that at least one species has a
substantial population size.
In this parameter range the model is not very sensitive to the
exact parameter values \cite{RIKV06}. 
We also note that the chosen connectance is above the percolation limit
on the $L=20$ dimensional cube of potential genotypes
\cite{GAVR04,STAU92}, so that a finite fraction of the genotypes can be 
connected by mutations along paths of nonzero interactions. 

The very long simulation times were chosen because our main
interest is in the statistically stationary
dynamics of macroevolution over timescales much
longer than the ecological ones of a few generations. 
Each run therefore starts with a ``warm-up period" of about one million
generations before the $2^{25}$-generation data-taking period. 
Details of the simulation algorithm were given in Ref.~\cite{RIKV03}. 

\subsection{Time series and species abundance distributions}
\label{sec:TS}

\begin{figure}[t] 
\begin{center}
\includegraphics[angle=0,width=.47\textwidth]{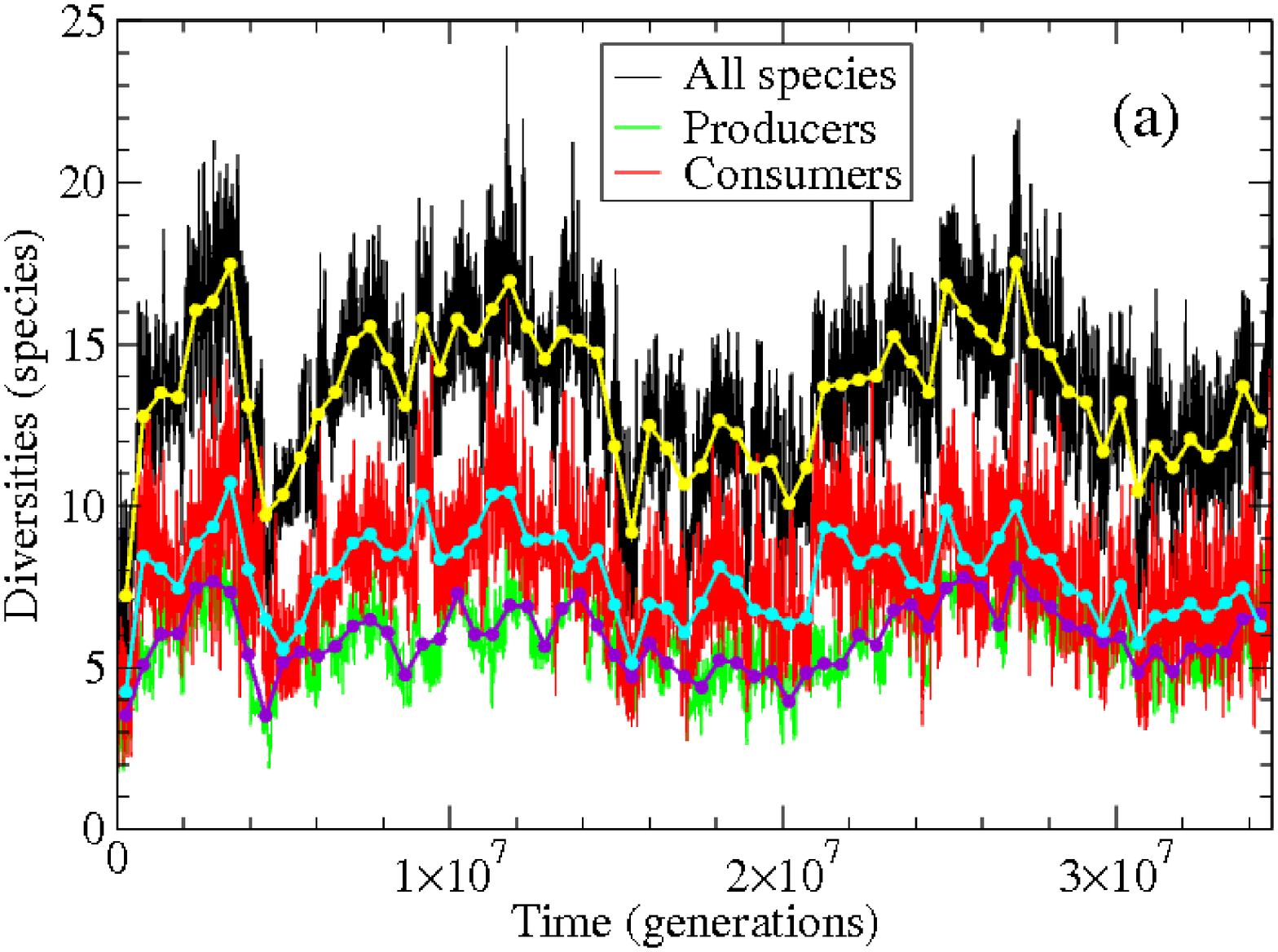}
\hspace{0.5truecm}
\includegraphics[angle=0,width=.47\textwidth]{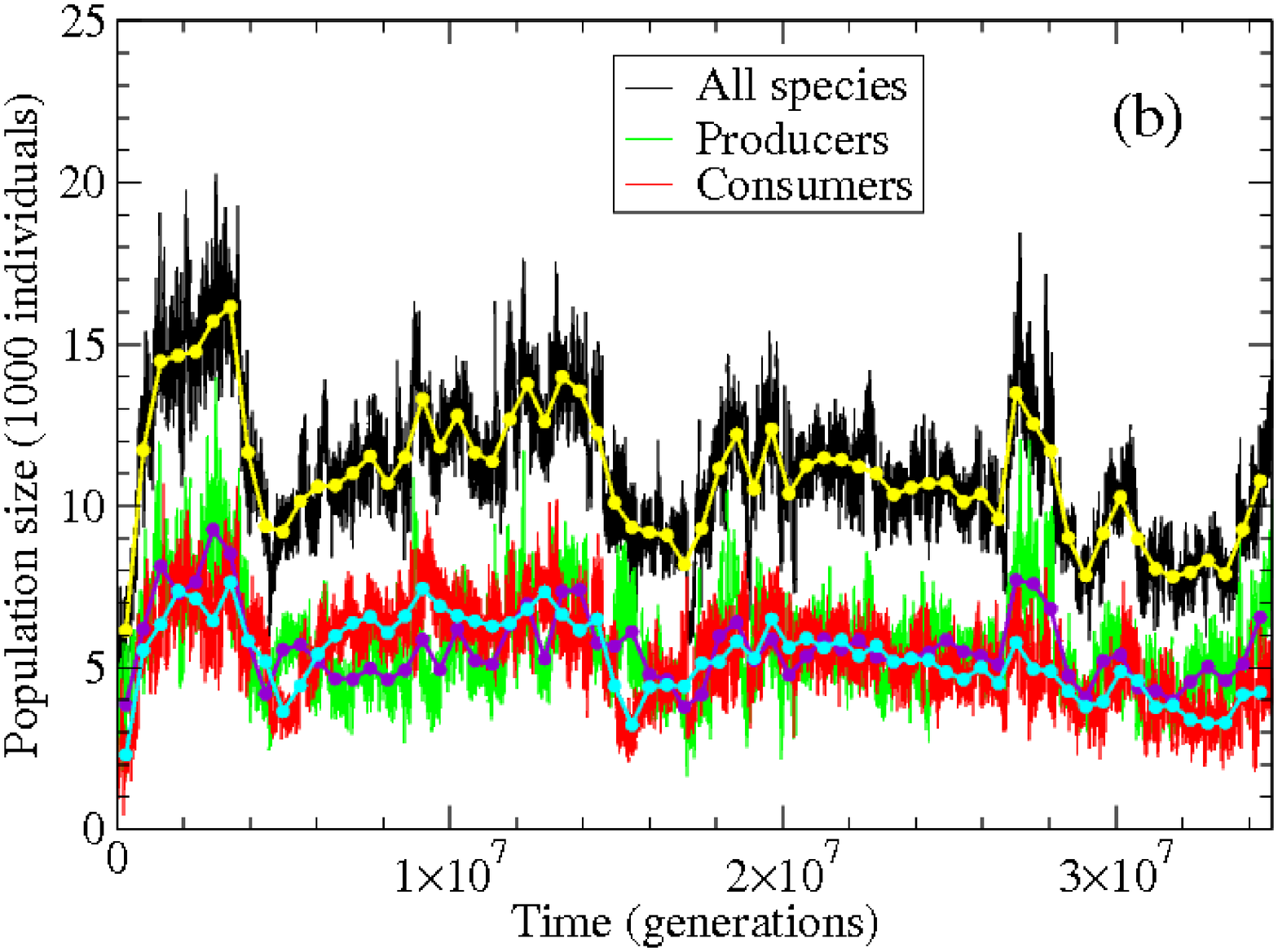} 
\end{center}
\caption[]{
(Color online.)
Time series for the Shannon-Wiener diversity index {\bf (a)} and the 
population sizes {\bf (b)}. The jagged curves in the background
show data sampled every 2048 generations to show the rapid
fluctuations, while the curves in contrasting color/brightness in
the foreground are running averages over 524\,288 generations,
emphasizing the slower fluctuations. The three sets of curves
represent all species (black/light gray; black/yellow online), 
producer species (light gray/dark gray; green/violet online), 
and consumer species (medium gray/light gray; red/cyan online). 
}
\label{fig:timser}
\end{figure}
We collected time series of a number of quantities including several
measures of diversity or species richness, as well as population 
sizes of producer and consumer species.
Time series of diversities and population sizes for 
one representative run are shown in Fig.~\ref{fig:timser}.

\begin{figure}[t]
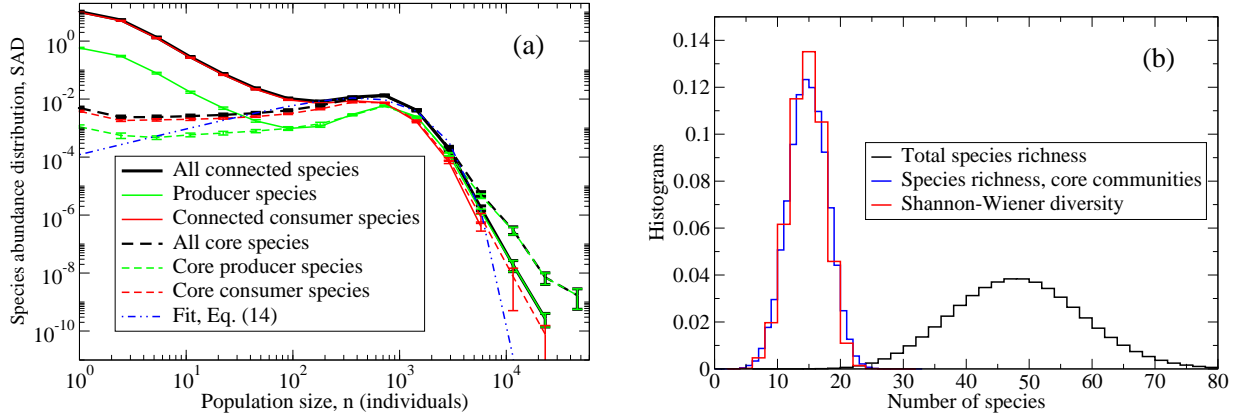
 
\begin{center}
\includegraphics[angle=0,width=.47\textwidth]{SADfigAllA.eps}
\hspace{0.5truecm}
\includegraphics[angle=0,width=.47\textwidth]{DivHistsALfig.eps} 
\end{center}
\caption[]{
(Color online.)
{\bf (a)}
Species abundance distributions (SADs), normalized to the number of
species. Solid curves represent ``full connected communities," and  
dashed curves represent ``core communities," both extracted as
described in the text. Both were sampled every 256 generations.
The data were averaged over twelve independent simulation runs, and the
error bars represent standard errors, based on the differences between
runs.  The dot-dashed curve is a fit to the curve describing all core species,
using Eq.~(\protect\ref{eq:PIGO}) with
parameters $C=13.569$, $\mu=0.00210967$, and $\beta=1.89338$,
interpolating between log-series and log-normal forms. 
{\bf (b)}
Histograms of the full-community species richness, the species richness of the
core communities, and the Shannon-Wiener diversity for the sampled
communities. The latter is seen
to be an excellent approximation for the species richness of the core
communities. 
}
\label{fig:SAD}
\end{figure}
\begin{figure}[t]
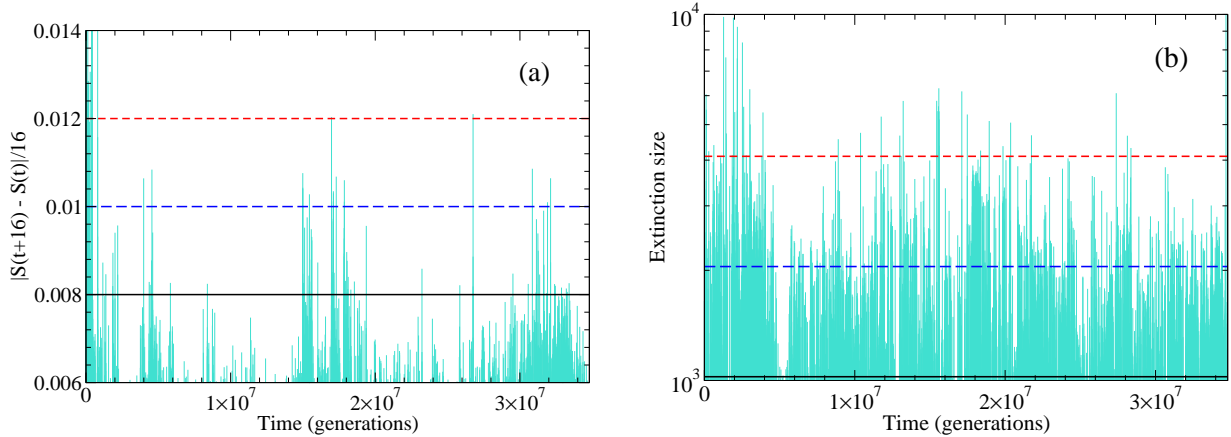
 
\begin{center}
\includegraphics[angle=0,width=.47\textwidth]{dSdTTimeSerFig.eps}
\hspace{0.5truecm}
\includegraphics[angle=0,width=.47\textwidth]{ExtTimeSer_256Fig.eps} 
\end{center}
\caption[]{
(Color online.)
Time series for the same simulation run shown in
Fig.~\protect\ref{fig:timser}, displaying  
quantities that measure the evolutionary activity: the magnitude of the
logarithmic derivative of the Shannon-Wiener diversity index, 
$|dS/dt|$
(averaged over 16 generations) {\bf (a)}, and the size of
extinctions per generation {\bf (b)}. The horizontal lines in each
part indicate cutoff levels used to define quiet and active
periods as discussed in Sec.~\protect\ref{sec:Q}.  
}
\label{fig:timser2}
\end{figure}

In order to filter out noise from 
the small-population species that are mostly
unsuccessful mutations, we use the diversity measure known in
ecology as the exponential Shannon-Wiener index \cite{KREB89}. 
It is defined as the exponential function of the information-theoretical
entropy of the population distributions, 
$D(t) = \exp \left[S \left( \{ n_I(t) \} \right) \right]$, where 
\begin{equation}
S\left( \{ n_I(t) \} \right)
=
- \sum_{\{I | \rho_I(t) > 0 \}} \rho_I(t) \ln \rho_I(t)
\label{eq:S}
\end{equation}
with
$\rho_I(t) = n_I(t) / N_{\rm tot}(t)$ for the case of the curves
labeled ``All species" in Fig.~\ref{fig:timser}. For the producers or
consumers, the sums and normalization constant include only the
appropriate species.
The utility of the Shannon-Wiener index is illustrated by the data
presented in Fig.~\ref{fig:SAD}. In Fig.~\ref{fig:SAD}(a) we show two
versions of the Species Abundance Distribution (SAD) for the model. This
is one of the tools most widely used in ecology to describe the
distribution of the number of species over population sizes \cite{HUGH86}.
The SADs shown as full lines in the figure represent full communities,
from which we have only removed any consumer species that are not
connected to the external resource through an unbroken chain of nonzero
interactions. These we term ``full connected communities" ("full
communities" for short). 
(The removal of disconnected consumer species actually has a numerically 
insignificant effect on the SADs.) 
The SADs for the full communities are dominated by
a large number of species with very small populations, which to a
large extent represent unsuccessful mutants. This was explicitly shown
by extracting ``core communities" in the following way. 
Communities were sampled every 256 generations, and 
species with population sizes below eight were excluded. 
It was then checked whether each included species also
existed with at least this minimum population 256 generations ago, and
if this was not the case, the species was removed from the community as
unstable. The fixed-point populations for the community of remaining
species were then calculated according to Eq.~(\ref{eq:ssn}), species
with negative fixed-point populations (unfeasible species) were removed,
and the fixed-point 
calculation was repeated until all species remaining in the community had
positive populations. The SAD was then calculated for each of these
feasible core communities and averaged over all communities in 
the twelve independent simulation
runs. This procedure removes most of the low-population
species, as shown by the dashed curves in
Fig.~\ref{fig:SAD}(a). The core-community SAD appears to be intermediate
between Fisher's log-series distribution \cite{HUGH86,FISH43} and Preston's
log-normal distribution \cite{HUGH86,PRES48}. It can be semiquantitatively
approximated by fitting the constants $C$, $\mu$, and $\beta$ in
the function \cite{PIGO04}
\begin{equation}
p(n) = C \mu^\beta n^{\beta -1} e^{-\mu n} / \Gamma(\beta) 
\;,
\label{eq:PIGO}
\end{equation}
which interpolates between these two limiting forms. 
The fit is shown by the dash-dotted curve in the figure. 
Additional evidence for the agreement between the Shannon-Wiener
diversity index and the species richness of the core communities is
shown in Fig.~\ref{fig:SAD}(b), where histograms of the two are in
excellent agreement and both show a narrow peak near fifteen species 
(mean values of 15.7 and 15.3 species, respectively), while the
raw species richness yields a wide distribution with a mean of 49.3 species. 
All three distributions are very well fit by gaussians and are much more
symmetric than diversity distributions arising from Rossberg et al.'s 
speciation model of food webs \cite{ROSS06A}. 

Time series of additional quantities that indicate the level of
evolutionary activity are shown in Fig.~\ref{fig:timser2} for the
same simulation run as in Fig.~\ref{fig:timser}. These are the magnitude 
of the logarithmic derivative of the Shannon-Wiener diversity index 
(Fig.~\ref{fig:timser2}(a)), and the size of extinctions per
generation (Fig.~\ref{fig:timser2}(b)). The latter is defined as
the sum of the maximum population sizes reached by the species that
go extinct in each generation.  

The properties of the fluctuations in the time series were analyzed
with several methods, including power spectral densities (PSD),
lifetimes of individual species, and the durations of quiet and active
periods during the evolution. The results of each are reported in the
following subsections. 

\subsection{Power spectral densities}
\label{sec:PSD}

\begin{figure}[t]
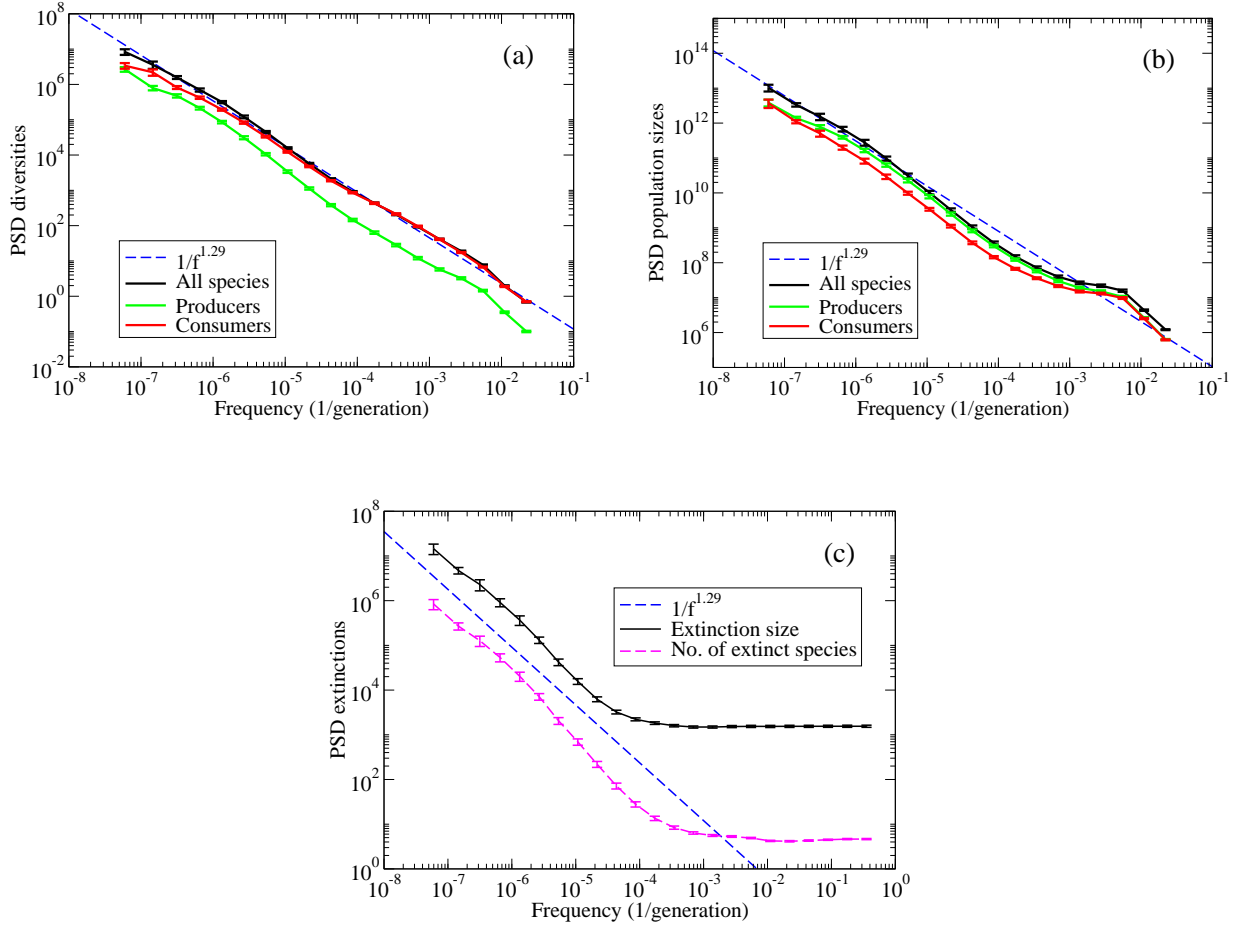
 
\begin{center}
\includegraphics[angle=0,width=.47\textwidth]{PSDfig_div.eps}
\hspace{0.5truecm}
\includegraphics[angle=0,width=.47\textwidth]{PSDfig_pop.eps} \\
\vspace*{1.0truecm}
\includegraphics[angle=0,width=.47\textwidth]{PSDfig_ext20.eps} 
\end{center}
\caption[]{
(Color online.)
PSDs for the Shannon-Wiener
diversities {\bf (a)}, population sizes {\bf (b)}, and
the extinction sizes and number of species going extinct per
generation {\bf (c)}, 
all averaged over twelve independent runs of $2^{25}$
generations. The error bars are standard errors,
estimated from the variations between the individual runs. The dashed straight
line with slope $-1.29$ in {\bf (a)} is a weighted fit to the PSD for the
overall diversity over the whole frequency range. 
The dashed lines in {\bf (b)} and {\bf (c)} are guides to the eye with 
the same slope.
}
\label{fig:PSD}
\end{figure}
PSDs of the diversity and population-size fluctuations, averaged over the
twelve independent simulation runs, are shown in Fig.~\ref{fig:PSD}, 
both for the total population and for the producers and 
consumers separately. 
The spectra indicate $1/f$ like noise over more than five decades
in time. A weighted fit to the PSD for the overall diversity in 
Fig.~\ref{fig:PSD}(a) yields a
power law $f^{- \alpha}$ with $\alpha \approx 1.29 \pm 0.01$.
This power is also seen to fit reasonably well, both with the data 
for the diversities of producers and consumers over the whole frequency
range in Fig.~\ref{fig:PSD}(a), and with the PSDs of all three
population measures at low frequencies in Fig.~\ref{fig:PSD}(b),
as well as for the extinction measures at low frequencies in 
Fig.~\ref{fig:PSD}(c). 
This suggests that the long-time fluctuations in the diversity, as
well as in 
the population sizes and the extinction measures, obey the same power law on 
long timescales. On short timescales the PSDs
for the population sizes have a more complicated structure, possibly
indicating overdamped oscillations on a scale of a few hundred
generations. The extinction measures show a wide region of white
noise for high frequencies, due to the frequent extinction of
unsuccessful mutants. However, the behaviors for low frequencies appear 
consistent with the diversities and the population sizes. 

\subsection{Species lifetimes}
\label{sec:life}

\begin{figure}[t] 
\begin{center}
\includegraphics[angle=0,width=.47\textwidth]{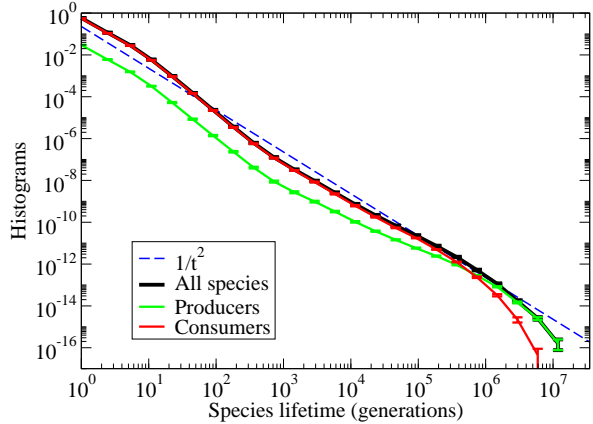}
\end{center}
\caption[]{
(Color online.)
Histograms for the lifetimes of individual species. The dashed, 
straight line is a guide to the eye, corresponding to a $t^{-2}$ power law. 
Data averaged over 12 independent simulation runs. 
}
\label{fig:SpecLife}
\end{figure}
The statistics of the lifetimes of individual species 
are characteristic of the evolution process.
The species lifetime is defined as the time from a particular 
species enters the community, till it goes extinct (i.e., its 
first return time to zero population size).
Histograms showing the distributions of species lifetimes, for all
species as well as for producers and consumers separately, are shown in
Fig.~\ref{fig:SpecLife}. Although there are some undulations in these
curves, they remain close to a power law $t^{-\tau_1}$
with exponent $\tau_1 \approx 2$ 
over more than six decades in time, which is the maximum we could expect with
the length of our simulations. 

The $t^{-2}$ dependence of the lifetime distributions is quite
universal. It is found, e.g., in our previous studies of the mutualistic model
\cite{SEVI06,RIKV05A,RIKV06}, and it is in general characteristic of stochastic
branching processes \cite{PIGO05}. 

Lifetime distributions for marine genera that are compatible
with a power-law exponent in the range $-1.5$ to $-2.5$
have been obtained from the fossil record \cite{NEWM99B,NEWM03,BORN06}.
However, the possible power-law behavior in the fossil record
is only observed over about one decade in
time -- between 10 and 100 million years -- and other fitting functions,
such as exponential or log-normal, are also possible.
Nevertheless, it is reasonable to conclude that the numerical results
obtained from complex, interacting evolution models that extend
over a large range of time scales support interpretations of the fossil
lifetime evidence in terms of nontrivial power laws.

\subsection{Quiet and active periods}
\label{sec:Q}

\begin{figure}[t]
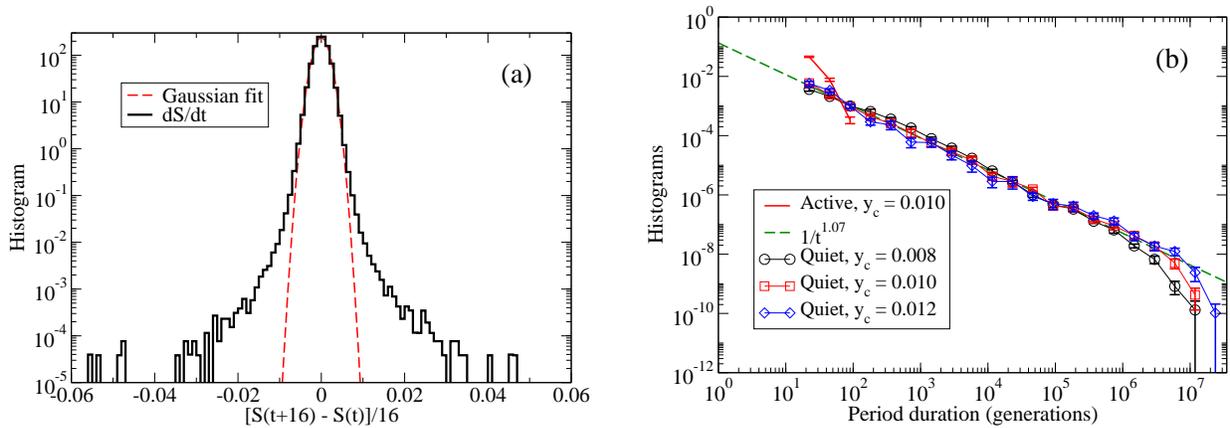

\begin{center}
\vspace{0.5truecm}
\includegraphics[angle=0,width=.47\textwidth]{dSdtHISTfig.eps}
\hspace{0.5truecm}
\includegraphics[angle=0,width=.47\textwidth]{DurQDSfig.eps} 
\end{center}
\caption[]{
{\bf (a)}
Histogram of the logarithmic derivative of the overall diversity,
$dS/dt$. The data were averaged over 16 generations for each run and
then over 12 independent runs. The dashed curve is a gaussian fit. 
The time series of the magnitude of
this quantity for one particular simulation run
is shown in Fig.~\protect\ref{fig:timser2}(a). 
{\bf (b)}
Histograms 
for the durations 
of active and quiet periods for various
values of the cutoff $y_c$ for $|dS/dt|$, averaged over 12 independent runs. 
The dashed, straight line with
slope $-1.07$ is a weighted fit to the histogram for the quiet periods
for $y_c=0.010$ between 10 and $10^6$ generations. 
The steep histogram 
curve with only three data points between 10 and 100 generations
corresponds to the active periods, which are always very short. 
}
\label{fig:dSdt}
\end{figure}
\begin{figure}[t]
\begin{center}
\includegraphics[angle=0,width=.47\textwidth]{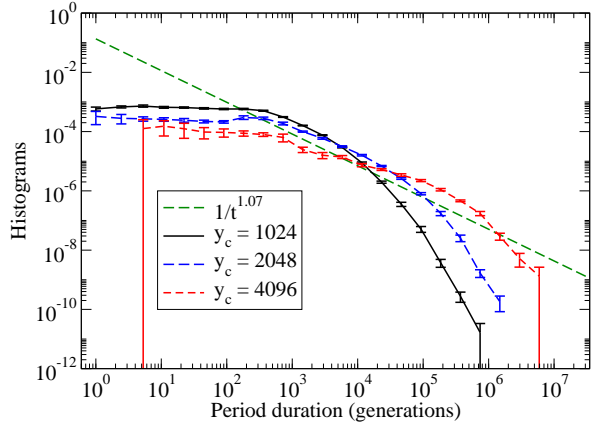}
\end{center}
\caption[]{
Histograms of the durations of quiet periods, obtained from the time
series of extinction sizes, an example of which is 
shown in Fig.~\protect\ref{fig:timser2}(b). 
Data averaged over 12 independent simulation runs. 
The straight, dashed line with slope $-1.07$ is a guide to the eye,
based on the fit to the QSS duration distribution based on $|dS/dt|$, 
shown in Fig.~\protect\ref{fig:dSdt}(b). 
}
\label{fig:QuietExt}
\end{figure}
{}From the time series shown in Figs.~\ref{fig:timser} and~\ref{fig:timser2} 
one sees that periods of moderate fluctuations are punctuated by periods of
high activity. The communities corresponding to the lower fluctuation
intensities are known as quasi-steady states (QSS).
(In the literature on the tangled-nature model \cite{CHRI02,HALL02,COLL03}
the QSS are referred to as quasi evolutionarily steady strategies, or q-ESS.) 
The cores of these communities correspond to the
fixed-point communities of the mutation-free system \cite{RIKV03,ZIA04}, 
as discussed in Sec.~\ref{sec:TS}. 
One measure of the degree of activity is the time
derivative of the entropy or, equivalently, the logarithmic
derivative of the total Shannon-Wiener 
diversity. Its magnitude is shown as a time series
in Fig.~\ref{fig:timser2}(a), and a histogram
is shown in Fig.~\ref{fig:dSdt}(a). While the central part of the
distribution is well approximated by a gaussian, the heavier wings
appear to be exponential or even power-law. Quiet and active periods were
defined as contiguous periods during which $|dS/dt|$ stayed below or
above a cutoff $y_c$, respectively. Octave-binned histograms for the
probability distributions of the durations of quiet and active periods
are shown in Fig.~\ref{fig:dSdt}(b) for various cutoffs. For the quiet 
periods a power law is
seen with exponent near $-1$ (a weighted fit between 10 and $10^6$
generations gives $t^{-\tau}$ with $\tau \approx 1.07\pm0.01$) and a long-time
cutoff that increases with increasing $y_c$. The active
periods for all values of $y_c$ are very brief in comparison. 
(Their histogram for $y_c = 0.010$ is the 
steep curve with only three data points between 10 and 100 generations
in Fig.~\ref{fig:dSdt}(b).) 
As a consequence, the system spends most of its time in QSS communities --
a situation consistent on the community level
with Eldredge and Gould's concept of punctuated
equilibria \cite{ELDR72,GOUL77,GOUL93}. 

Durations of quiet periods could also be obtained from the
time series of extinction sizes in Fig.~\ref{fig:timser2}(b). Due
to the white noise at short timescales, which was also apparent in
the PSDs in Fig.~\ref{fig:PSD}(c), the power-law behavior is
limited to a window of longer times between about 1000 generations
and the strongly cutoff-dependent long-time decay. As a result, this
quantity does not provide as clear a quiet-period
distribution as the entropy derivative. We therefore did not
perform any independent fit to obtain a power-law exponent for the
QSS durations measured this way. See Fig.~\ref{fig:QuietExt}.  
The decay with time is qualitatively consistent with that observed 
in Fig.~\ref{fig:dSdt}(b) for the QSS duration distributions based on
$|dS/dt|$. 

It is quite remarkable that the exponent
for the QSS durations is significantly different from the one for
the species lifetimes, which is close to 2. This is particularly
so because the two exponents appear to be approximately the same (both near 2)
for the mutualistic version of the model \cite{SEVI06,RIKV05A,RIKV06}.
We believe that the explanation
lies in the structure of the QSS communities generated by the present evolution
process, which take the form of simple food webs. These are studied
in Sec.~\ref{sec:web} below.

\subsection{QSS community structure and stability}
\label{sec:web}

\subsubsection{General considerations}
\label{sec:webA}

\begin{figure}[t] 
\begin{center}
\includegraphics[angle=0,width=.47\textwidth]{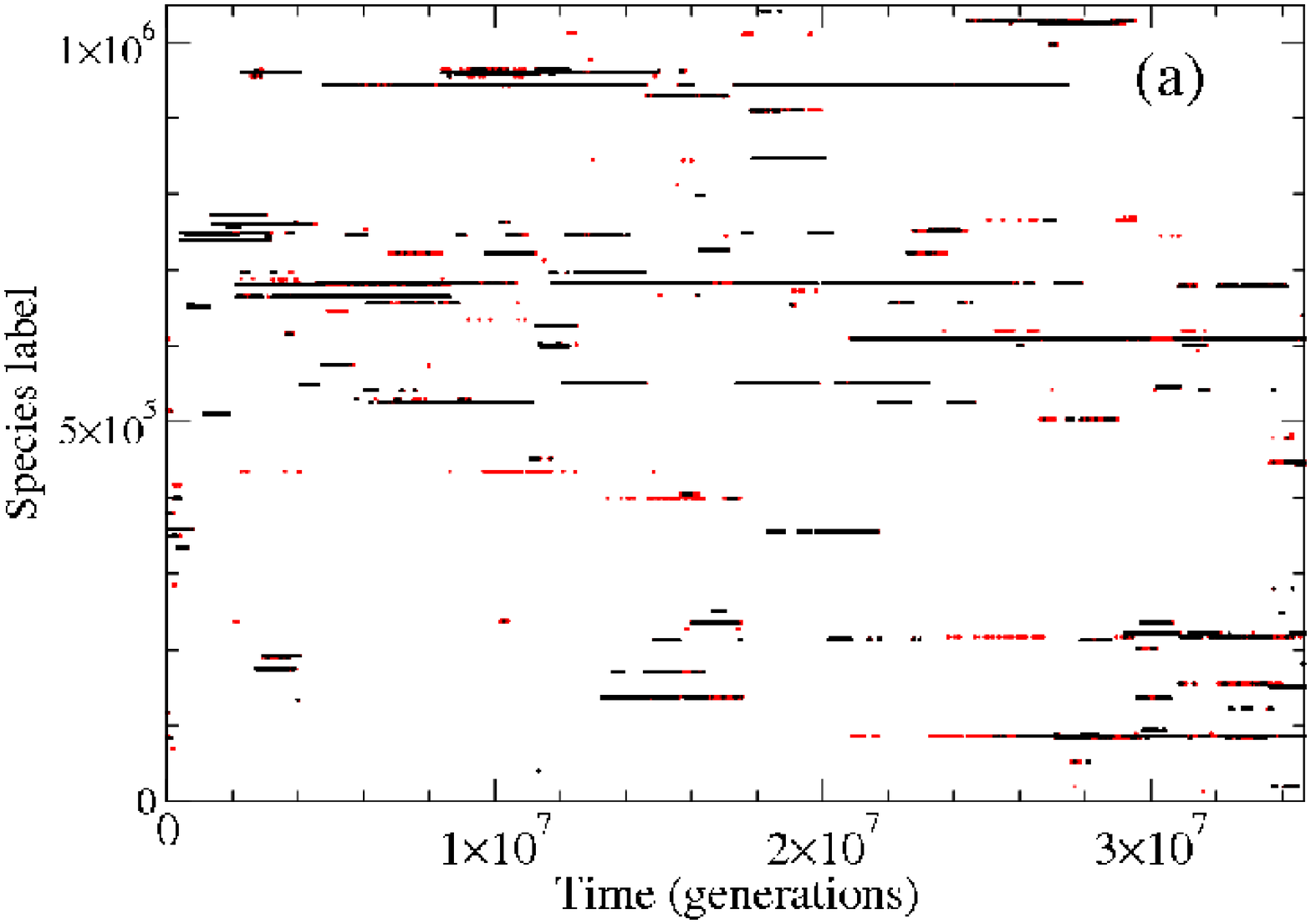}
\hspace{0.5truecm}
\includegraphics[angle=0,width=.47\textwidth]{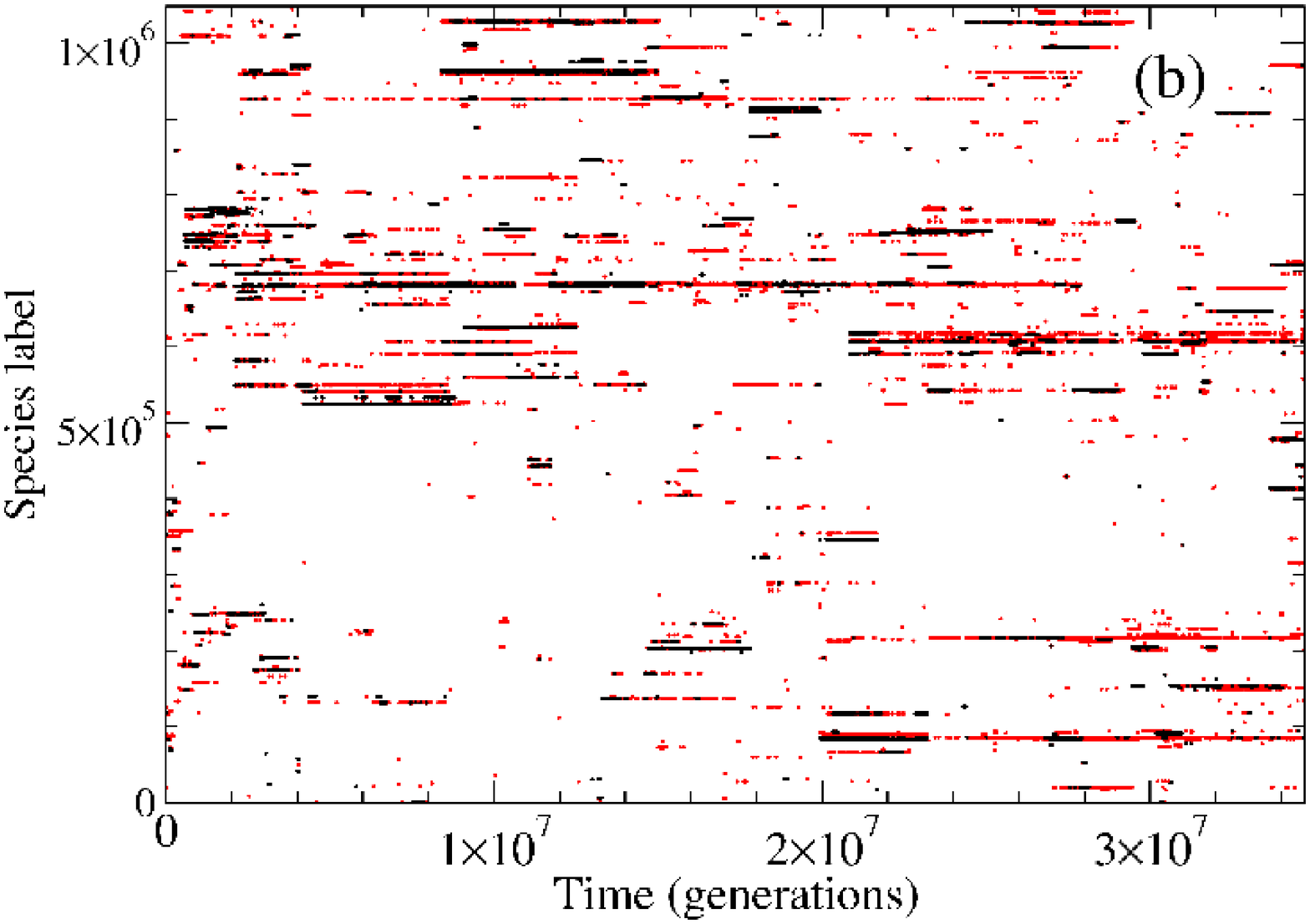}
\end{center}
\caption[]{
(Color online.)
Major populated species shown vs time for the same simulation run shown in 
Figs.~\protect\ref{fig:timser} and~\protect\ref{fig:timser2}.
The horizontal lines correspond to the
species label, and the grayscale (color online)
to the population size. Black: $n_I >
1000$. Gray (red online): $n_I \in [101,1000]$. 
{\bf (a)}: Producers. 
{\bf (b)}: Consumers. 
}
\label{fig:species}
\end{figure}
The evolution process in the present model generates dynamic
communities, in which species emerge, exist for a shorter or longer
time, and eventually go extinct. The emergence and extinction of a major
species are quite fast processes on the evolutionary timescale, and so
the vast majority of randomly selected communities are QSS communities. 
This is confirmed by the short durations of evolutionarily active
periods, shown by the corresponding histogram in Fig.~\ref{fig:dSdt}(b). 
Diagrams of the population sizes of
major producer and consumer species as functions of time in a particular
simulation run are shown in Figs.~\ref{fig:species}(a)
and (b), respectively. In these figures a horizontal line represents
a species. The beginning of the line represents the
emergence of the species, and the end 
represents its extinction. The population size is represented
by the color. We see that some species persist for tens of
millions of generations, while others are so short-lived as to hardly be
visible on the scale of these figures. This is consistent with
the power-law behavior of the species-lifetime distribution (see
Fig.~\ref{fig:SpecLife}). We also see that producer
species appear to emerge and go extinct relatively independently of
each other, while there is a significant correlation between the
producer and consumer species. 
This correlation indicates that extinction of a producer species is
likely to trigger a (limited) cascade of consumer extinctions.
Conversely, a new producer species is likely to quickly acquire a
group of consumer species. 
The structure of the plots in Fig.~\ref{fig:species} contrasts
with that of similar plots for the mutualistic model of Ref.~\cite{RIKV03} 
(see Fig.~2 of that paper),
in which species tend to emerge as well as go extinct together.
We believe this is the reason for the difference between the exponents for
the species lifetimes and the durations of QSS communities in this
model: the overall community is relatively resilient toward losing or
gaining a single species \cite{DROS01B,DUNN02,DUNN04,QUIN05B}. 
As a community it is more long-lived than the
individual species, leading to the smaller value for the exponent of the
distribution for the QSS durations (compare Figs.~\ref{fig:SpecLife}
and~\ref{fig:dSdt}(b)). (A similar relationship has been noted between
the lifetimes of orders and their constituent genera in the fossil
record \cite{BORN06}.)

\begin{figure}[t] 
\begin{center}
\includegraphics[angle=0,width=.47\textwidth]{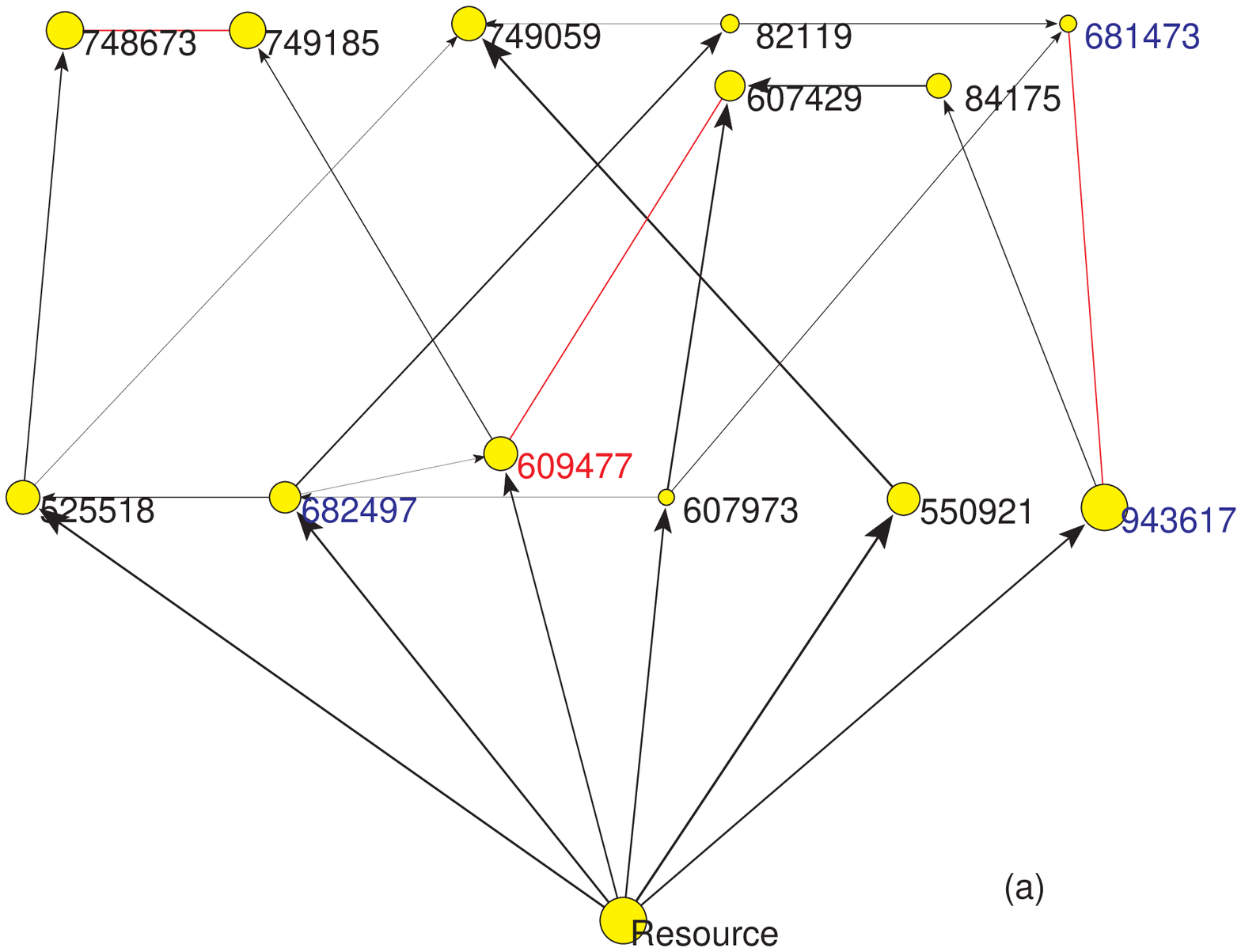}
\hspace{0.5truecm}
\includegraphics[angle=0,width=.47\textwidth]{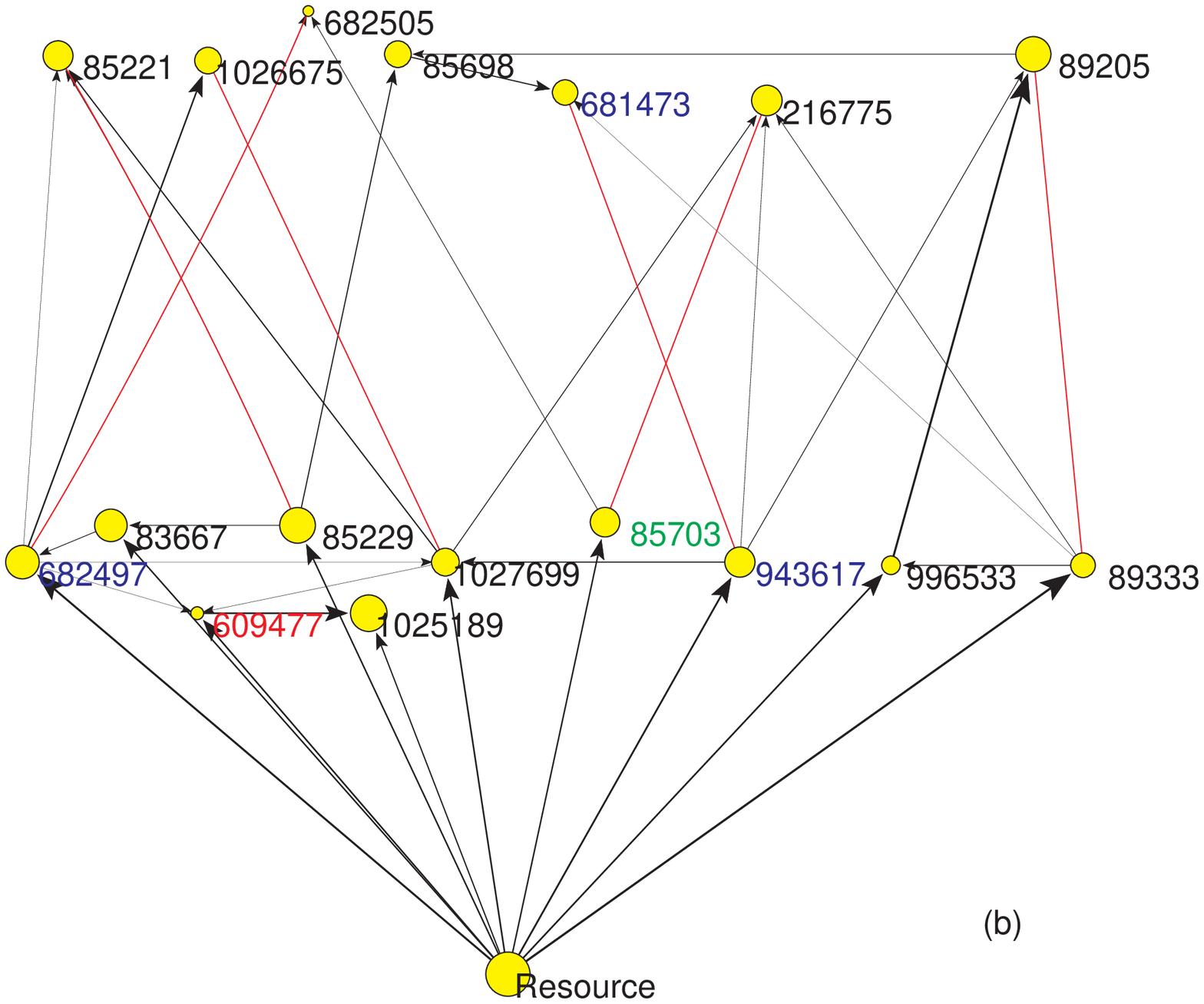} \\
\vspace{0.5truecm}
\includegraphics[angle=0,width=.47\textwidth]{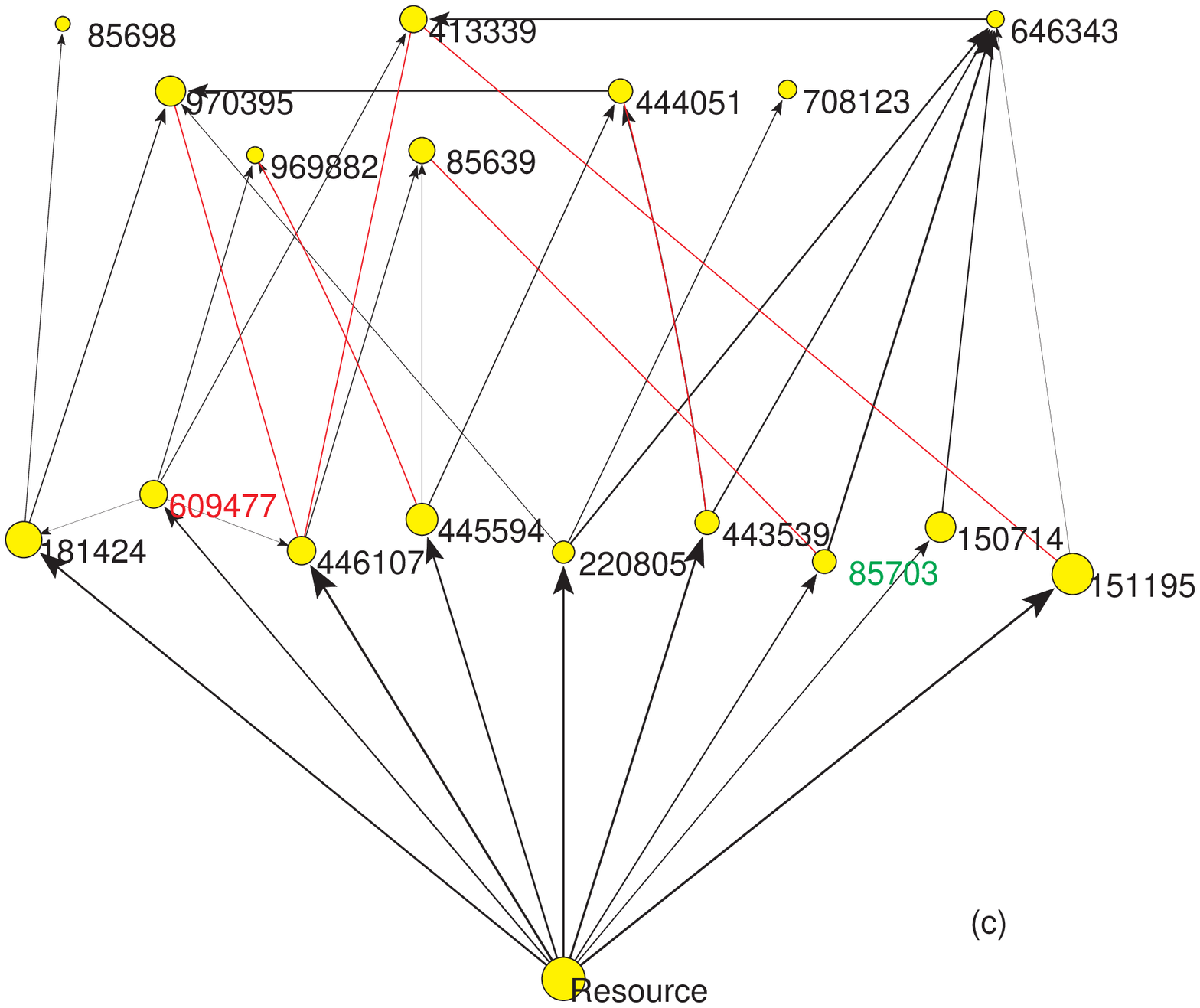} \\
\end{center}
\caption[]{
(Color online.)
Food webs representing QSS core communities 
for the simulation shown in Figs.~\protect\ref{fig:timser},
\protect\ref{fig:timser2}, and~\protect\ref{fig:species} at times 
near $22 \times 10^6$ generations {\bf (a)}, 
near $27 \times 10^6$ generations {\bf (b)}, 
and at the end of the simulation near 
$35 \times 10^6$ generations {\bf (c)}. 
The core communities were identified as described in
Sec.~\protect\ref{sec:TS}. 
The thickness and head size of the arrows correspond to the
magnitude of $M_{IJ}$, and the area of the circles to the
stationary population size, as calculated analytically from
Eq.~(\ref{eq:ssn}). Light gray (red online) 
lines connect nearest neighbors in genotype space. 
Labels in red mark species that persist from the first to the last
snapshot, {\bf (a)} to {\bf (c)},
while blue labels mark species that persist from {\bf (a)} to 
{\bf (b)}, and and green labels mark species that persist from {\bf (b)} 
to {\bf (c)}. Black labels indicate species that are unique to the
particular community.  
}
\label{fig:webs}
\end{figure}
Three representative QSS core communities from the same simulation run shown in
Figs.~\ref{fig:timser}, \ref{fig:timser2}, and~\ref{fig:species} 
are shown in Fig.~\ref{fig:webs}. These are consecutive core
communities near $22 \times 10^6$, $27 \times 10^6$, and $34 \times
10^6$ generations, respectively, and they are all stable by the eigenvalue
criterion discussed in Sec.~\ref{sec:stab}. 
The communities take the form of simple
food webs with two trophic levels above the resource node. Communities
with three trophic levels are also occasionally observed. The different
branches of the webs are relatively independent of each other, 
and many consumer species have more than one prey. 
(See further quantitative discussion in Sec.~\ref{sec:webB}.) Both features 
contribute to the resilience against mass extinctions discussed above. 

\begin{figure}[t] 
\begin{center}
\includegraphics[angle=0,width=.47\textwidth]{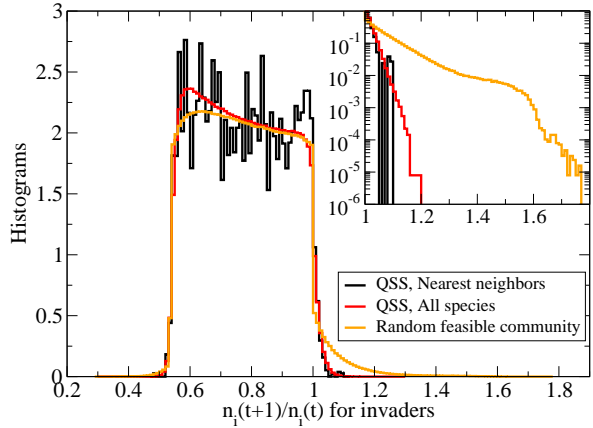}
\end{center}
\caption[]{
(Color online.)
Histograms of the multiplication ratio, $n_i(t+1)/n_i(t)$ (exponential
of the invasion fitness, Eq.~(\protect\ref{eq:inv})) 
for low-density invaders. Each curve is 
averaged over 14 different core communities. Black: nearest
neighbors in genotype space against stable QSS communities. 
Dark gray (red online): All species not in the community against stable
QSS communities.
Light gray (orange online): All species not in the community against
random feasible communities. 
Inset:
The part of the distributions for potentially successful invaders, 
$n_i(t+1)/n_i(t) > 1$, with the
$y$-axis on logarithmic scale. The tails appear nearly exponential. 
}
\label{fig:invader}
\end{figure}
The relative stability of evolved QSS core communities with respect to
invasion by mutants is illustrated in Fig.~\ref{fig:invader}. The
results are averaged over 14 QSS core communities -- the three shown in
Fig.~\ref{fig:webs} plus the final communities of the eleven other
simulation runs. This
figure shows the multiplication ratio of a small population of invaders 
(the exponential function of the invasion fitness),
given by Eq.~(\ref{eq:inv}). Only about 2.3\% of species outside the
communities have multiplication ratios larger than unity, and most of
these lie between 1.0 and 1.1. This
percentage does not seem to depend significantly on the Hamming distance
of the invader from the community. 
We note that the species that are removed from the community during
extraction of the core community are among these low invasion fitness
species. This is a further indication that the difference between core and
full communities is mostly made up by unsuccessful mutants. 

We also tested if the evolved core communities are more
resilient toward invasion than randomly constructed feasible communities.
Such communities are more difficult to construct in this model, than in
the mutualistic model \cite{RIKV03}. To have the
same level of statistics as for the QSS communities, we produced 14 such
communities in the following way. 
We started a run with a random sample of 200 species, each with
$n_I=10$, and evolved the community for 1024 generations {\it without\/}
mutations. We then tested the remaining community for feasibility and
removed species with a negative stationary population according to
Eq.~(\ref{eq:ssn}). The resulting feasible 
communities had much smaller diversities
than the evolved QSS communities -- an average of only 3.6 species per
community. These communities had a significantly larger percentage of
potential invaders -- about 4\%, with a maximum multiplication ratio
near 1.7 (see inset in Fig.~\ref{fig:invader}). While there thus is a
clear difference between the stability against invaders of QSS
communities and random feasible communities, the difference is not as
large as it is for the mutualistic model of Ref.~\cite{RIKV03}. (See Fig.~3
of that paper.)  

\subsubsection{Detailed community structure and comparison with real
food webs}
\label{sec:webB}

\begin{figure}[t] 
\begin{center}
\includegraphics[angle=0,width=.47\textwidth]{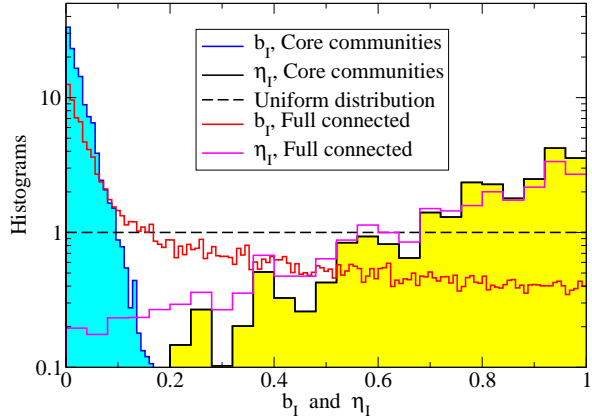}
\end{center}
\caption[]{
(Color online.)
Log-linear plot of 
histograms of the birth cost $b_I$ (left) and producer coupling to the 
external resource, $\eta_I$ (right), for members of QSS core communities 
(shaded areas) and full communities (lines only). 
Both parameters are selected away from the uniform distributions of
the full species pool (represented by the horizontal dashed line). 
}
\label{fig:b-eta}
\end{figure}
Next we turn to a detailed statistical description of the QSS core
and full connected communities identified in Sec.~\ref{sec:TS}.
Since communities were extracted every 256 generations from twelve
independent runs of $2^{25}$ generations each, data were averaged over 
$131\,072 \times 12$ communities of each type. 
However, due to the long-time correlations in the
evolution process, many of the communities from the same run
are identical or similar. This sampling
method thus ensures that the statistics for both communities and individual
species are weighted according to their longevities.
First we consider the properties of individual species, 
and next we turn to the collective
properties of the corresponding food webs and a comparison with data for
real food webs. 

\begin{figure}[t]
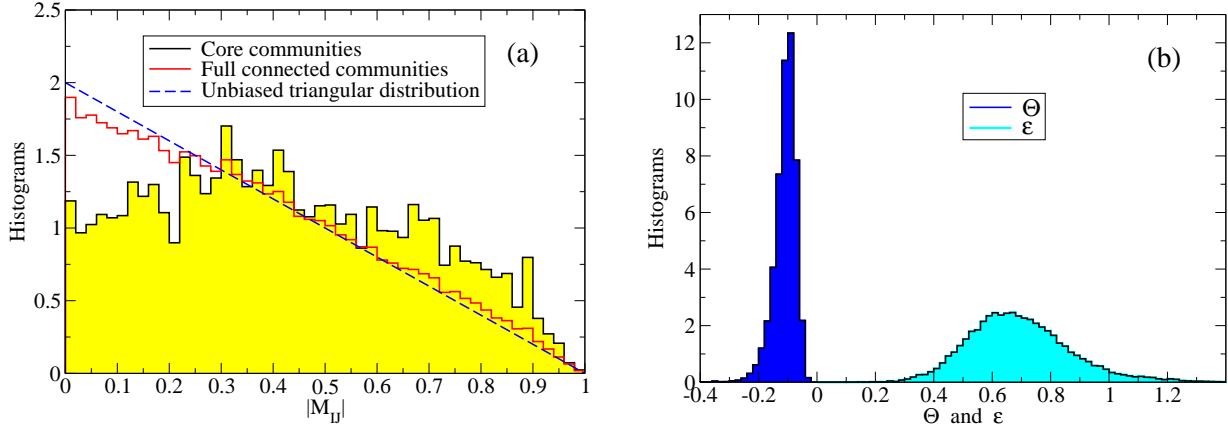
 
\begin{center}
\includegraphics[angle=0,width=.47\textwidth]{PosMijALfig.eps}
\hspace{0.5truecm}
\includegraphics[angle=0,width=.47\textwidth]{ThetaEfig.eps} 
\end{center}
\caption[]{
(Color online.)
Histograms of community-level parameters.
{\bf (a)}
The interspecies interaction strengths, $|M_{IJ}|$, for QSS core and
full communities. 
{\bf (b)}
The effective interaction strength $\Theta$ (left) and resource coupling
$\mathcal{E}$ (right) for QSS core communities, 
defined in Eq.~(\protect\ref{eq:quad2}). 
}
\label{fig:Mij}
\end{figure}
The individual species are characterized by the birth cost $b_I$, the
self interaction $M_{II}$, and for producers by the resource coupling
$\eta_I$. In the total pool of potential species these are uniformly
distributed on $(0,+1]$, $[-1,0)$, and $(0,+1]$, respectively. 
Not surprisingly, the most prevalent 
species in the core communities turn out to
be the most ``individually fit" ones in the sense that they have low
birth cost and, for producers, relatively strong coupling to the
external resource. The resulting probability densities for $b_I$ and
$\eta_I$ are shown in Fig.~\ref{fig:b-eta}. 
The selection for low birth cost is very strong for members of the core
communities, and less so when the full communities are considered. This
is further indication that the species that are ignored when extracting
the core communities have higher $b_I$, and thus overall are less
individually fit, than the core species. 
On the other hand, the moderate selection for strong coupling to the external
resource (large $\eta_I$) is approximately equal for producers in both
types of communities. In contrast, the self
interactions appear to be evolutionarily neutral, and their 
probability density remains approximately uniform for the members 
of both core and full communities (not shown).  

The first community-level quantities we consider are the nonzero
interspecies interaction strengths, $|M_{IJ}|$. Histograms for these,
based on the same sampling and averaging as previous results, are shown
in Fig.~\ref{fig:Mij}(a). The distribution for the core communities is
significantly skewed toward strong interactions, compared to
the triangular distribution characteristic of the total species pool,
which is shown as a dashed line. The bias is much weaker when all
members of the full communities are considered. 
(See Appendix~\ref{sec:AA} for a discussion of the unbiased distribution.)

\begin{figure}[t] 
\begin{center}
\vspace{0.5truecm}
\includegraphics[angle=0,width=.47\textwidth]{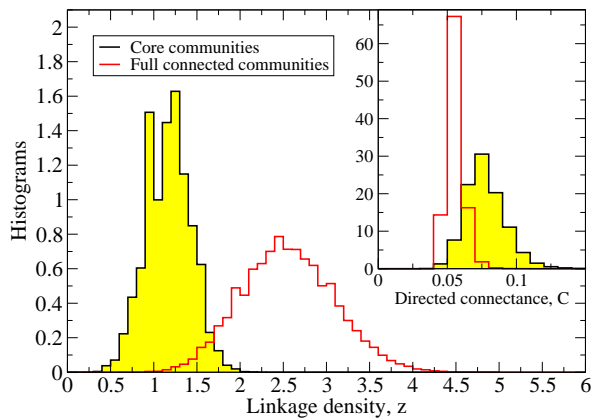}
\end{center}
\caption[]{
(Color online.)
Histograms for the linkage density, $z=L/S$ (main part) and
the directed connectance, $C=L/S^2$ (inset), for the model communities. 
}
\label{fig:conn}
\end{figure}
The biased distribution of interaction strengths combines with the strongly
skewed distributions of $b_I$ and $\eta_I$ to produce values of 
the effective interaction strength $\Theta$ and effective resource coupling
$\mathcal{E}$ for the core communities
(defined in Eq.~(\protect\ref{eq:quad2})) that are in
excellent agreement with the expectations expressed in
Sec.~\ref{sec:fpc}. As shown in Fig.~\ref{fig:Mij}(b), 
$\Theta$ is narrowly distributed closely below 0 ($\overline{\Theta} =
-0.10$), while $\mathcal{E}$ has a broader distribution over positive
values with mean $\overline{\mathcal{E}} = 0.71$.  
These averages are in excellent agreement with those obtained from the
final communities of the twelve runs, obtained in Sec.~\ref{sec:TS}. 

Next we consider several quantities from network theory
that can be compared with data from
real food webs. These include the directed connectance, $C=L/S^2$, where $L$
is the number of links (i.e., the number of 
nonzero $M_{IJ}$, $M_{JI}$ pairs) and $S$ is the diversity
(here defined as the species richness),
the linkage density, $z=L/S=CS$, the probability
distributions of a species' number of prey species (its generality or
indegree), number of predator species (its vulnerability or outdegree),
and their sum (its total degree) \cite{ROSS06A}. 

For comparison with the simulated food webs generated by our evolutionary
model, we use data for seventeen empirical webs, including both aquatic
and terrestrial communities. These data sets 
were kindly provided to us by J.~A.\ Dunne. 
Tabular overviews of the parameters characterizing these
communities are available in the literature  
\cite{DUNN02,DUNN02B,DUNN04,QUIN05}. For all the empirical
webs we defined the diversity as the species richness $S$ in terms
of {\it trophic species\/}. These
are obtained by lumping together all taxa that share all their predators
and prey \cite{WILL00,DUNN02B}. Like the simulated communities analyzed, the
empirical webs contain no disconnected species or subwebs \cite{DUNN02B}. 
The included empirical food webs are: 
Coachella Valley \cite{Polis:1991}, 
El Verde Rainforest \cite{Waide:1996}, 
Scotch Broom \cite{Memmott:2000}, 
St.\ Martin Island \cite{Goldwasser:1993}, 
and 
U.K.\ Grassland \cite{Martinez:99} 
(terrestrial);
Bridge Brook Lake \cite{Havens:1992}, 
Little Rock Lake \cite{Martinez:1991},
and 
Skipwith Pond \cite{Warren:1989} 
(lake or pond); 
Canton Creek \cite{Townsend:1998} 
and 
Stony Stream \cite{Townsend:1998}
(stream);
Chesapeake Bay \cite{Baird:1989}, 
St.\ Mark's Estuary \cite{Christian:1999}, 
Ythan Estuary \cite{Hall:1991},
and 
Ythan Estuary with parasites \cite{Huxham:1996},
(estuaries); 
Benguela \cite{Yodzis:1998},
Caribbean Reef -- small \cite{Opitz:1996}, 
and 
North-eastern U.S.\ continental shelf  \cite{Link:2002}
(marine).

As discussed above, the 
food webs produced by the evolutionary process in this model are
relatively small, with average diversity $\overline{S} \approx 15$ for
the QSS core communities and approximately 50 for the full communities. 
As seen in
Fig.~\ref{fig:conn}, the directed connectance $C$ is somewhat smaller
($\overline{C} \approx 0.08$ for core and 
$\overline{C} \approx 0.06$ for full communities) 
than the input connectance, $c=0.1$. However,
this is partly because it is calculated in the conventional way as $L/S^2$ 
\cite{ROSS06A}, rather than as $L/S(S-1)$. The average linkage density is
also quite small ($\overline{z} = 1.17$ for core and $\overline{z} =
2.58$ for full communities) compared to most of the real
food webs that have been documented. For comparison, the seventeen
documented food webs have $\overline{S} \approx 69$ with a range from
25 to 155,
$\overline{C} \approx 0.13$ with a range from 0.03 to 0.32, 
and $\overline{z} \approx 6.9$ with a range from 1.6 to 17.8. 
However, Williams and Martinez'
niche model for food-web structure \cite{WILL00} leads to a scaling
hypothesis for the degree distributions 
(exact for the niche-model prey distribution and approximate for the
predator distribution) \cite{CAMA02A,CAMA02B,STOU05}:
\begin{equation}
2z p(k) = \tilde{p}_k(k/2z) \;,
\label{eq:scale}
\end{equation}
where $k$ can be either the generality, the vulnerability, or the total
degree (with different scaling functions $\tilde{p}$
for each). As a consequence,
cumulative degree distributions can also be rescaled by simply dividing
the argument by $2z$. Using this niche-model result as a general scaling
hypothesis \cite{ROSS06A}, we can compare the degree distributions for the
communities generated by our model with scaled data for the seventeen 
real food webs. 

\begin{figure}[t]
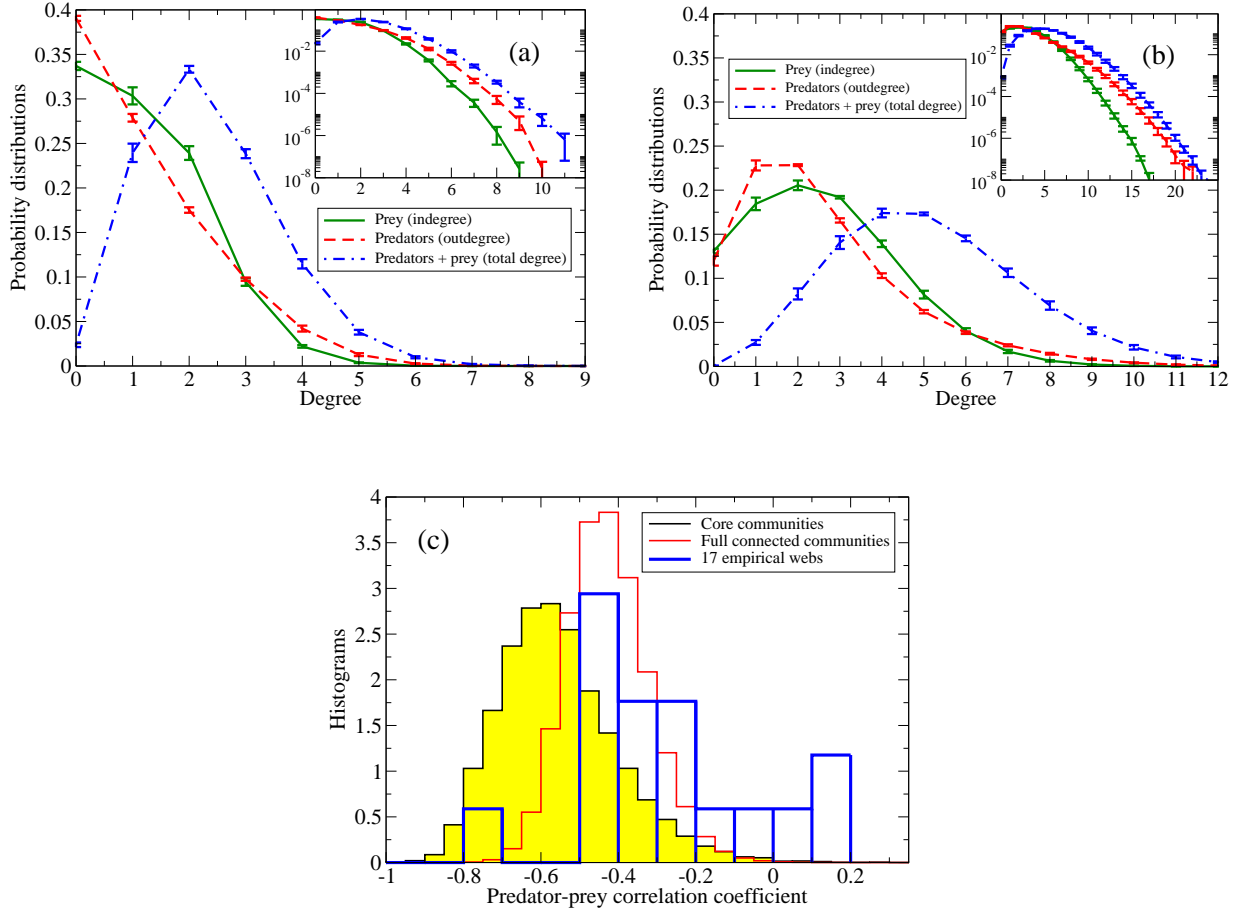
 
\begin{center}
\includegraphics[angle=0,width=.47\textwidth]{Pkfig.eps}
\hspace{0.5truecm}
\includegraphics[angle=0,width=.47\textwidth]{Pkfig_full.eps}\\
\vspace{1.0truecm}
\includegraphics[angle=0,width=.47\textwidth]{Corfig.eps} 
\end{center}
\caption[]{
(Color online.)
{\bf (a)}
Unscaled degree distributions for the QSS core communities, 
showing number of prey (solid),
number of predators (dashed), and total degree (dot-dashed). 
The distributions were individually normalized for each community,
and then averaged over all communities. 
The error bars are standard errors, based on the spread
between the twelve simulation runs.
Average linkage density is $\overline{z} = 1.17$.
The inset on log-linear scale shows that these
distributions decay at least exponentially with increasing argument.  
{\bf (b)}
Same as (a), for the full communities. 
Average linkage density is $\overline{z} = 2.58$.
{\bf (c)}
Histograms over individual sampled communities of the 
correlation coefficient between a species' numbers of predators and
prey. Black filled with light gray (yellow online): QSS core communities. 
Medium gray (red online): full, connected communities. 
Dark gray (blue online) with drop lines: empirical data for seventeen natural
food webs. 
}
\label{fig:Pk}
\end{figure}
The degree distributions for the model, individually normalized for each
community, and then averaged over all communities, are shown in unscaled
form in Fig.~\ref{fig:Pk}(a) and~(b). As seen from the insets, 
all three distributions (prey,
predators, and total degree) decay at least exponentially for large
argument. This property is shared by most 
real food webs \cite{DUNN02B} 
and models \cite{CAMA02A,CAMA02B,ROSS06A,SEVI06B} 
and indicates that food webs in general are not scale-free networks. 
The behavior for $k < 2z$ is more model and system dependent. 

Histograms for 
the correlations between a species' generality and vulnerability are 
shown in Fig.~\ref{fig:Pk}(c), together with data for the seventeen real food
webs. The average correlation coefficients are $-0.53$ for core communities and 
$-0.40$ for full communities, more negative than,
but not inconsistent with, the real food-web average of $-0.23$.  

\begin{figure}[t]
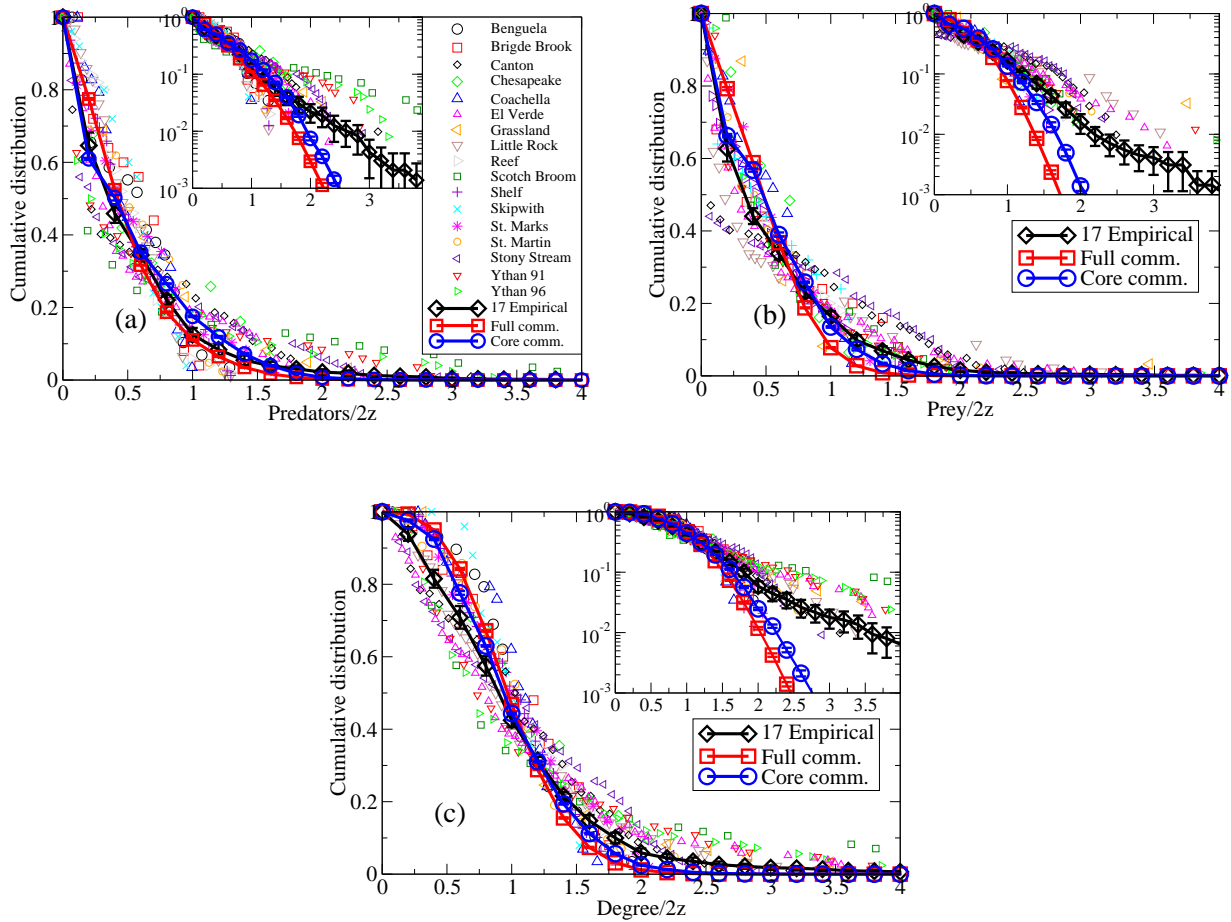
 
\begin{center}
\includegraphics[angle=0,width=.47\textwidth]{Cum_Pred_Scale_fig.eps}
\hspace{0.5truecm}
\includegraphics[angle=0,width=.47\textwidth]{Cum_Prey_Scale_fig.eps}\\
\vspace{1.0truecm}
\includegraphics[angle=0,width=.47\textwidth]{Cum_Total_Scale_fig.eps} 
\end{center}
\caption[]{
(Color online.)
Scaled, cumulative degree distributions for the model, 
compared with results for the seventeen empirical food webs. 
Data for individual, real food webs are shown in the background as
isolated data points. Average data for the simulated and real webs are
shown as data points with error bars indicating standard error,
connected by heavy lines. 
Dark gray with circles (blue online): simulated core communities. 
Medium gray with squares (red online): simulated full communities. 
Black with diamonds: empirical communities. 
Details of the averaging are discussed in the text. 
{\bf (a)}
Number of predators (vulnerability, or outdegree). 
{\bf (b)}
Number of prey (generality, or indegree). 
{\bf (c)}
Number of predators plus prey (total degree). 
}
\label{fig:Cum}
\end{figure}
Due to the relatively small size of the empirical food webs, it is
difficult to extract information from scaled probability densities,
which will include empty bins unless a very large bin size is chosen.
This problem is avoided by instead studying 
the cumulative distribution, $P(x) = {\rm Prob}\left[k/(2z) > x\right]$ 
\cite{DUNN02B}. These are shown in Fig.~\ref{fig:Cum}
for the seventeen empirical webs, together with averaged results for the
empirical webs and the model. The averaged curves were generated by
first scaling the degrees (predators, prey, and total) for each web by 
the value of $2z$ for that particular web, binning the results in bins
of width 0.2, and normalizing the binned histogram by the species
richness $S$ of that same web. The individually scaled, binned, and
normalized histograms
were then averaged over all webs (seventeen for the empirical webs and 
$131\,072 \times 12$ for the model webs) and finally integrated to produce the
average scaled cumulative distributions. For the model, results are shown both
for the full and core communities. 

Despite their smaller average diversity and linkage density, the 
scaled cumulative degree distributions for both types of model
communities, which are
shown in Fig.~\ref{fig:Cum}, fall well within the range of
the empirical data. For scaled degrees below about 1.5, the averaged
distributions for the core communities are in quite reasonable
agreement with those for the empirical food webs. The largest deviations
are for total degree (Fig.~\ref{fig:Cum}(c)) at small values of the
scaled argument. This is possibly due to the significantly stronger
negative correlations between the numbers of predators and prey
(vulnerability and generality) for species in the model communities,
compared to the empirical ones. (See Fig.~\ref{fig:Pk}(c).) 
For scaled degrees above approximately 1.5, the averages for the
empirical webs are significantly above those for the model webs and appear
approximately exponential in the tail (see insets in
Fig.~\ref{fig:Cum}). We note, however, that this behavior in the
averages is caused by a small number of webs; the remaining
empirical webs have no species of such highly above-average degrees. 
Overall, the scaled, cumulative distributions for the present model
agree with the empirical data to about the same degree as the niche
model \cite{STOU05} and the speciation model of Rossberg et al.
\cite{ROSS05,ROSS06A}. However, it must be admitted that much
potentially valuable detail about the food-web structure is
lost in calculating the cumulative distributions. 

\begin{figure}[t]
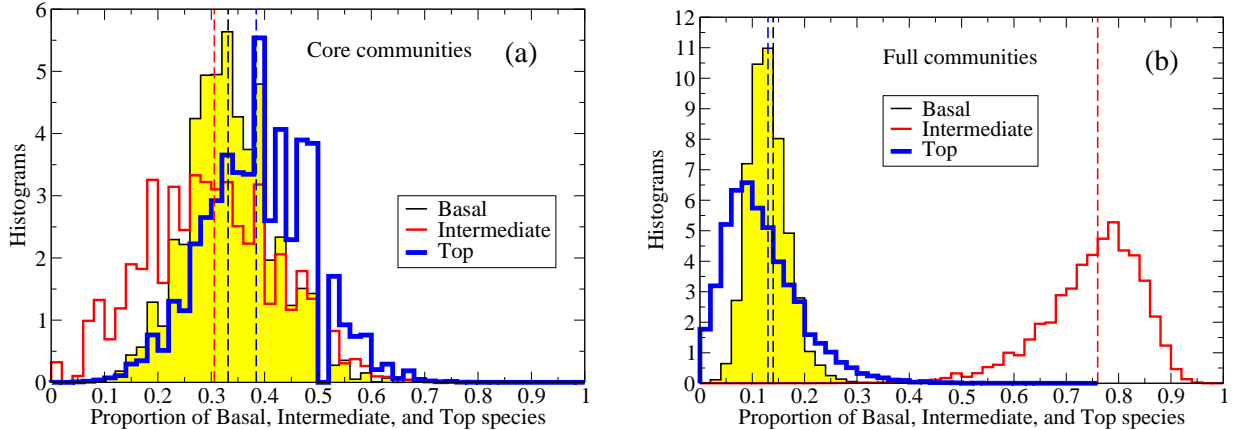
 
\begin{center}
\includegraphics[angle=0,width=.47\textwidth]{BITfig.eps}
\hspace{0.5truecm}
\includegraphics[angle=0,width=.47\textwidth]{BITfig_full.eps}
\end{center}
\caption[]{
(Color online.)
Histograms for the proportions of basal, intermediate, and top species.
The vertical, dashed lines in matching grayscale (color) represent the
mean proportion for each class. 
{\bf (a)}
Core communities. Mean proportions: 0.33 for basal, 0.31 for intermediate,
and 0.38 for top species.
{\bf (b)}
Full communities. Mean proportions: 0.14 for basal, 0.76 for intermediate,
and 0.13 for top species.
}
\label{fig:BIT}
\end{figure}

A measure of the hierarchical structure of a food web is given by the
proportions of basal species (species with no prey, supported only by the
external resource), intermediate species (that have both prey and
predators) and top species (that have no predators). As seen from the
histograms in Fig.~\ref{fig:BIT}, there are wide variations from
community to community for each
class of species. The high proportion $I$ of intermediate species seen for
the full communities in Fig.~\ref{fig:BIT}(b) ($I=0.76$) is consistent with
most of the real food webs analyzed, including 
Caribbean Reef ($I=0.94$),
U.S.\ Shelf ($I=0.94$),
Benguela ($I=0.93$),
Scotch Broom ($I=0.92$), 
Skipwith Pond ($I=0.92$), 
Coachella Valley ($I=0.90$), 
Little Rock Lake ($I=0.86$), 
El Verde Rainforest ($I=0.69$), 
St.\ Marks Seagrass ($I=0.69$), 
St.\ Martin Island ($I=0.69$), 
and 
Bridge Brook Lake ($I=0.68$),
as well as simulations of the Web World model \cite{QUIN05} and
mean-field analysis of a Lotka-Volterra predator-prey 
model with evolution and competition \cite{LASS01}.  The core communities,
on the other hand, have their member species much more evenly distributed among
the three classes, as shown in Fig.~\ref{fig:BIT}(a). 
Real food webs with more even distributions are 
Canton Creek ($B=0.53$, $I=0.22$, $T=0.25$),
Stony Stream ($B=0.56$, $I=0.27$, $T=0.17$), 
and 
Chesapeake Bay ($B=0.16$, $I=0.52$, $T=0.32$). 
Thus, at least in this model, and somewhat counterintuitively,
on average the intermediate species are both the least stable and  
the most numerous. 

We conclude these comparisons of the structures of food webs evolving in
the present model to data for real food webs with some words of
caution. The model was primarily developed to study the long-time temporal
fluctuations in diversity and population size, and it was in no way
tuned to produce realistic community structures except insofar as they
are generally food-web like. This community structure does, however have
profound effects on the dynamics, as it is responsible for the
difference between  
the power-law exponents for the lifetimes of species and communities in
this predator-prey model. 
One should also note that, while the lack of correlations between the
properties of closely related species was shown to have little effect on
the long-time dynamics \cite{SEVI06}, we have not considered whether
such correlations might affect the detailed community structure. Most
likely they would, at least to some degree.
In addition, the growth of real food webs involve a number of
mechanisms, besides just evolution and local population dynamics. Most
notably, immigration plays an important role in the occurrence of new
members of a spatially localized community. However, due to the large
differences between parent species and mutants in the current model, our
evolution mechanism may also been seen as covering immigration. 

We also note that the simulated communities have significantly smaller
diversities, connectances, linkage densities, and population sizes than
the real communities, so that our comparisons had to rely heavily on
scaling arguments. It would therefore be desirable in future studies to
increase the connectance, number of potential species, and resource size
to bring the results closer to the realistic range.

\section{Summary and Conclusions}
\label{sec:Concl}

In this paper we have studied in detail an individual-based
predator-prey model of biological coevolution, based on the simplified
version of the tangled-nature model \cite{CHRI02,HALL02,COLL03}
that was introduced in Ref.~\cite{RIKV03}. Selection is provided by a
population-dynamics model in which the reproduction probability of an
individual of a particular species depends nonlinearly on the amount of
external resources and on the population densities of all other species
resident in the community. New species appear in the community through
point mutations in a genome consisting of a string of $L$ bits. 

In the mutation-free limit, the mean fixed-point population sizes
and stability properties of any $\mathcal{N}$-species community can be 
obtained exactly by linear stability analysis. While the universal competition
effect that enables this analytical treatment is not very realistic, the
exact solutions make the model ideal as a benchmark for more realistic,
but also more complicated, models. A preliminary discussion of two more
realistic models is found in Ref.~\cite{RIKV06C}. 

In the simulations presented here, we used $L=20$ for a total of $2^{20} =
1\,048\,576$ potential species. In order to study the statistically
stationary properties of the model, we performed long kinetic Monte
Carlo simulations over $2^{25} = 33\,554\,432$ generations. 
By studying the stationary fluctuations we hope in the future to gain an
understanding of the the system's sensitivity to external perturbations
in a way analogous to the fluctuation-dissipation relations of
equilibrium statistical mechanics \cite{SATO03}. 

Qualitatively, many of the statistical properties of this model are
similar to those of the related, mutualistic model studied in
Refs.~\cite{RIKV03,ZIA04,SEVI06}. These include approximate $1/f$ noise
in power spectra (PSDs) of diversity and population sizes 
($f^{- \alpha}$ with $\alpha \approx 1.29$), and
power-law distributions for the lifetimes of individual species, as well
as of the durations of evolutionarily quiet periods, corresponding to
QSS communities. However, in contrast to the mutualistic model, the
power-law exponents for the species lifetimes and QSS durations are
different: $t^{-\tau_1}$ with $\tau_1 \approx 2$ for the former 
(consistent with a stochastic branching process \cite{PIGO05}) 
and $t^{-\tau}$ with $\tau \approx -1.07$
for the latter. In Ref.~\cite{RIKV05A} it was speculated that the
exponent values $\alpha=1$, $\tau_1=2$, and $\tau=1$ are consistent with
predictions for a zero-dimensional extremal-dynamics
model \cite{PACZ96,DORO00}. 
However, this speculation is not consistent with the presumably more
accurate estimate for $\alpha$ presented here, and it seems advisable  
to be rather skeptical about any mapping of the current model onto a
simple statistical-mechanical extremal-dynamics model. 

It is probably more fruitful to consider why $\tau_1$ and $\tau$ are different
for the current predator-prey model, while they coincide for the
corresponding mutualistic model. Here we believe the answer lies in the
different community structures in the two models. While communities in
the mutualistic model are tightly knit with all positive interactions
(see Fig.~10 of Ref.~\cite{RIKV06}), the communities in the present
predator-prey model take the form of simple food webs, which are much more
resilient toward the loss of a single or a few species. In this sense,
this predator-prey model is in much better agreement with real food webs
than the mutualistic model \cite{DROS01B,DUNN02,DUNN04,QUIN05B}. 

Our comparisons of the structure of the communities generated by our
model with real food webs show both similarities and differences. A
difficulty with such comparisons is the large differences in diversity
and linkage density between the simulated and real webs, which
necessitated heavy use of scaling arguments in the quantitative
comparisons. For comparison of detailed community properties, 
simulations of models with higher diversity and connectance
are therefore desirable in the future. 
Nevertheless, our model presents a synthesis of long-term
evolutionary dynamics and food-web like communities within an
individual-based framework of integrated population dynamics and
evolution, that should provide a sound basis for more refined models in
the future.

\section*{Acknowledgments}
\label{sec:ACK}

We are grateful to J.~A.\ Dunne for providing us with raw data
on seventeen empirical food webs. 
We also thank two anonymous referees for helpful comments on an earlier
version of this paper. 

This research was supported in part by National Science Foundation Grant 
Nos.~DMR-0240078 and DMR-0444051, and by Florida State 
University through the School of Computational Science,
the Center for Materials Research and Technology, the National High
Magnetic Field Laboratory, and a COFRS summer salary grant.

\appendix

\section{Matrix Elements for Larger Genomes}
\label{sec:AA}

Here we describe an improved version of the method introduced by
Hall et al. \cite{HALL02,CHRI02} to produce
pseudorandom matrix elements $M_{IJ}$ for values of $L$ that are too large 
for the full $2^L \times 2^L$ matrix
$\bf M$ to fit into computer memory. This method
permits the matrix elements for a given community to
be generated and retained only as needed. 
We first present Hall et al.'s method, point
out some problematic features, and then present our modifications. 

Let ${\bf S}(I)$ be the string of binary digits corresponding to the 
decimal species label $I$. This bit string has length $L$, so there are
$2^L$ different strings. To generate the matrix element $M_{IJ}$, one
first generates a new string of the same length, 
${\bf S}(I,J) = {\bf S}(I) \, {\sf XOR} \, {\bf S}(J)$, where ${\sf XOR}$ 
is the logical {\it exclusive or\/} operator. 
From this bit string is generated the corresponding new decimal index 
$K\left( {\bf S}(I,J) \right)$. 
Next one creates two one-dimensional arrays, ${\tt X}$ and ${\tt Y}$, 
each of $2^L$ random numbers between $-1$ and $+1$. (For simplicity let 
the starting index for the arrays be zero.) Since ${\bf S}(I,J)$ is
symmetric in $I$ and $J$, asymmetric pseudorandom matrix elements 
are generated as 
\begin{equation}
M_{IJ} = \left[ {\tt X}\left( K\left( {\bf S}(I,J) \right) \right) 
+ {\tt Y} \left( J \right) \right] / 2 \;.
\label{eq:mjensen}
\end{equation}
(Hall et al. instead use the product of the two random numbers, which
gives a pseudorandom number with a slightly different distribution.)

\begin{figure}[t]
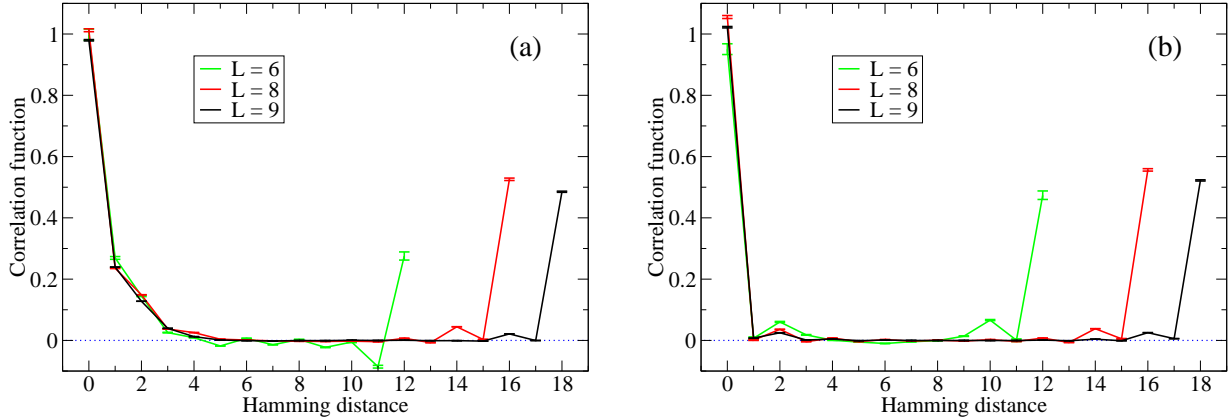
 
\begin{center}
\includegraphics[angle=0,width=.47\textwidth]{Mij_correl_Jensen_fig.eps} 
\hspace{0.5truecm}
\includegraphics[angle=0,width=.47\textwidth]{Mij_correl_X_fig.eps} 
\end{center}
\caption[]{
(Color online.)
Normalized
correlation functions for the matrix elements $M_{IJ}$ for the two
schemes described in the text, each 
based on a single realization of $\bf M$ for $L=6$, 8, and 9. 
{\bf (a)}
The scheme described in Ref.~\protect\cite{HALL02}. 
{\bf (b)}
The modified scheme introduced here. 
}
\label{fig:corr}
\end{figure}
The problem with this method is that it produces strong correlations
along the columns of $\bf M$ since the second of the two random numbers
that produce $M_{IJ}$ is the same for all $I$ at the same $J$. In
Fig.~\ref{fig:corr}(a) we show the resulting correlation function for this
scheme as a function of the Hamming distance between the pairs of bit
strings involved in two different matrix elements. 
Regardless of $L$, significant correlations are seen for Hamming
distances less than five. A discussion of how to calculate such
correlation functions is found in Ref.~\cite{SEVI06}.  

\begin{figure}[t] 
\begin{center}
\includegraphics[angle=0,width=.47\textwidth]{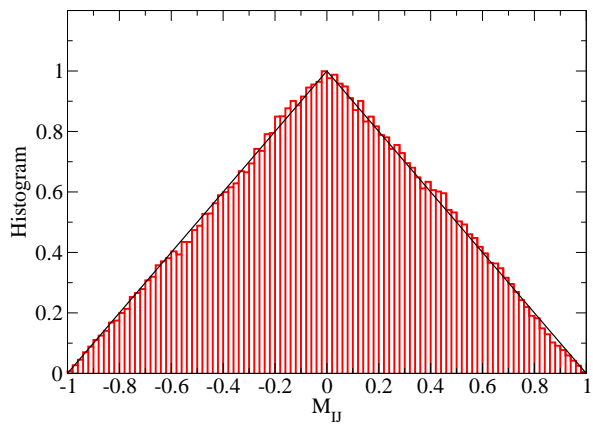} 
\end{center}
\caption[]{
(Color online.) 
Histogram for the pseudorandom $M_{IJ}$ produced by the scheme
introduced here for $L=9$, corresponding to 262\,144 individual matrix
elements. The straight, black lines represent the triangular shape of the
theoretical distribution.  
}
\label{fig:dist}
\end{figure}
To reduce these correlations between matrix elements involving closely
related genotypes, we here modify the scheme as follows. 
We extend array $\tt Y$ to a length of $3 \times 2^L$ and define
$M_{IJ}$ as 
\begin{equation}
M_{IJ} = \left[ {\tt X}\left( K\left( {\bf S}(I,J) \right) \right) 
+ {\tt Y} \left( K\left( {\bf S}(I,J) \right) + 2(J+1) \right) \right] / 2 \;.
\label{eq:X}
\end{equation}
The addition of $K\left( {\bf S}(I,J) \right)$ in the index of $\tt
Y$ ensures that the matrix element depends in an erratic fashion on both
$I$ and $J$, while the term linear in $J$ ensures that $\bf M$ is not
symmetric. The resulting correlation function is shown in 
Fig.~\ref{fig:corr}(b). The correlations for elements involving closely
related genotypes are strongly suppressed. The correlations for
a Hamming distance of $2L$, which are present in both schemes,
are of little practical significance. They are caused by the fact
that the ${\sf XOR}$ operation is invariant under simultaneous
bit reversal in both its arguments and could be removed by adding a
linear function of $I$ in the argument of $\tt X$. 
The probability density for $M_{IJ}$ is triangular as expected from
simple analytical arguments, and the whole interval from $-1$ to $+1$ is
well sampled, even for relatively small $L$. This is shown in 
Fig.~\ref{fig:dist}. 




\end{document}